\documentclass[seceq]{ptptex}
\usepackage{wrapft}
\usepackage{graphicx}
\newcommand{\bra}[1]{\langle {#1} |}     %%
\newcommand{\ket}[1]{| {#1} \rangle}     %%
\newcommand{\wtilde}[1]{\widetilde{#1}} %%

\newcommand{\gsim}{{\stackrel{\displaystyle >}{\raisebox{-1ex}{$\sim$}}}}

\def\beq{\begin{eqnarray}}
\def\eeq{\end{eqnarray}}
\def\bsub{\begin{subequations}}
\def\esub{\end{subequations}}
\def\b{\begin{equation}}
\def\bs{\begin{split}}
\def\es{\end{split}}
\def\e{\end{equation}}
\title{Interplay between Spin Polarization and Color Superconductivity 
in High Density Quark Matter}

\author{%       %Use \sc for the family name
Yasuhiko {\sc Tsue},$^{1,2}$ 
Jo\~ao da {\sc Provid\^encia},$^{3}$ 
Constan\c{c}a {\sc Provid\^encia},$^{3}$\\ 
Masatoshi {\sc Yamamura}$^{4}$ and Henrik {\sc Bohr}$^{5}$
}

\inst{%         %Affiliation, neglected when [addenda] or [errata]
$^{1}$Centro de F\'isica Computacional, Departamento de F\'{i}sica, Universidade de Coimbra, 3004-516 Coimbra, 
Portugal\\
$^{2}$Physics Division, Faculty of Science, Kochi University, Kochi 780-8520, Japan\\
$^{3}$Departamento de F\'{i}sica, Universidade de Coimbra, 3004-516 Coimbra, 
Portugal\\
$^{4}$Department of Pure and Applied Physics, 
Faculty of Engineering Science, Kansai University, Suita 564-8680, Japan\\
$^{5}$Department of Physics, B.307, Danish Technical University, DK-2800 Lyngby, Denmark
}

\abst{
Here, it is suggested
that a four-point interaction of the tensor type may lead to spin
polarization in quark matter at high density. It is found that the
two-flavor superconducting phase and the spin polarized phase correspond
to distinct local minima of a certain generalized thermodynamical
potential. 
It follows that
the transition from one to the other phase occurs passing through true minima 
with both a spin polarization and a color superconducting gap.
It is shown that the quark spin polarized phase is realized at rather high density, while 
the two-flavor color superconducting phase is realized at a lower density region.
}

\begin{document}
\maketitle

\section{Introduction}

Under extreme conditions, such as high temperature and/or high baryon density, 
it is interesting to study the behavior of 
quark and gluonic matter and/or hadronic matter governed by quantum chromodynamics (QCD) 
in the context of the physics of the relativistic heavy ion collisions  
and of the interior of compact stars. 
At low temperature and high density region \cite{FH} such as the interior of a neutron star, 
it is widely investigated and pointed out that various phases may be realized, for example, 
various meson condensed phases in hadronic matter \cite{Tamagaki}, the two-flavor color superconducting (2SC) and 
the color-flavor locked (CFL) phases \cite{RW,IB,CS}, the quarkyonic phase \cite{MP} 
and the quark ferromagnetic phase \cite{Tatsumi} in quark matter, 
and so forth. 
Especially, we may conjecture that the quark ferromagnetic phase exists at high density quark matter, 
since the existence of magnetar was reported \cite{Magnetar}. 
In a preceding work \cite{Tatsumi}, the pseudovector-type interaction between quarks was considered and it was pointed out that 
the quark spin polarization was realized. 
However, it was shown that, the quark spin alignment disappears, if the quark mass is zero, for example, 
in the chiral symmetric phase \cite{Medan}. 
Thus, it is interesting to investigate whether there is a possibility of spontaneous 
spin polarization under another interaction between quarks.

Recently, the present authors have indicated that the quark spin polarization may occur at high density 
quark matter even in the chiral symmetric phase 
due to the tensor-type four-point interaction between quarks, which leads to the quark spin polarization \cite{1,arXiv}. 
In that paper, it has been shown that a second-order phase transition occurs from normal quark phase to quark spin polarized phase. 
However, in quark matter at high baryon density, it is believed that the two-flavor color superconducting phase appears. 
Thus, it is interesting to investigate the stability of the quark spin polarized phase against the 2SC phase. 

In this paper, we investigate the color-symmetric superconducting phase \cite{2} and the quark spin polarized phase and clarify which phase is stable 
by using the Nambu-Jona-Lasinio (NJL) model \cite{NJL} including quark-pair interaction \cite{Kitazawa} 
and the tensor-type four-point interaction \cite{1,arXiv}. 
The quark-pair interaction is derived by the Fierz transformation from the original NJL model Lagrangian. 
As for the color superconducting phase, many possibilities, such as the gapless superconducting phase 
which is still controversial for realistic situations, have been considered.
%As for the color superconducting phase, many possibilities such as the gapless 
%superconducting one, and so on, 
%are considered and it may be still controversial for realistic situations. 
However, in this paper, the simple 2SC phase derived by the NJL model is treated. 
We thus consider the quark-pairing gap $\Delta$ and the quark spin polarization $F$ such that 
$\Delta\neq 0$ and/or $F\neq 0$ leads to the 2SC phase and/or quark spin polarized phase, respectively. 
In this paper, the system is treated by the mean field approximation and the 
BCS state is introduced \cite{3}. 

This paper is organized as follows: 
In the next section, the basic Hamiltonian 
is introduced and expressions of the Hamiltonian under some basis sets are given. 
In Appendix A, one of the expressions of the Hamiltonian is given in which 
good helicity states are used. 
In \S 3, the BCS state is defined, where detail derivation is given in Appendix B, 
and the thermodynamic potential is derived. 
In \S 4, numerical results and discussions are given together with Appendix C 
and the last section is devoted to a summary 
and concluding remarks.

\section{NJL model with tensor interaction}

\subsection{Basic Hamiltonian}

Let us start with the following Lagrangian density: 
\beq\label{1}
& &{\cal L}={\cal L}_0+{\cal L}_S+{\cal L}_T+{\cal L}_c\ , \nonumber\\
& &\ {\cal L}_0={\bar \psi}i\gamma^\mu\partial_\mu\psi\ , \nonumber\\
& &\ {\cal L}_S=G_S(({\bar \psi}\psi)^2+({\bar \psi}i\gamma_5{\vec \tau}\psi)^2)\ , 
\nonumber\\
& &\ {\cal L}_T=-\frac{G}{4}\left(({\bar \psi}\gamma^{\mu}\gamma^{\nu}{\vec \tau}\psi)
({\bar \psi}\gamma_{\mu}\gamma_{\nu}{\vec \tau}\psi)
+({\bar \psi}i\gamma_5\gamma^{\mu}\gamma^{\nu}\psi)
({\bar \psi}i\gamma_5\gamma_{\mu}\gamma_{\nu}\psi)\right)\ , \nonumber\\
& &\ {\cal L}_c=\frac{G_c}{2}\sum_{A=2,5,7}\left(
({\bar \psi}i\gamma_5\tau_2\lambda_A\psi^C)(
{\bar \psi}^Ci\gamma_5\tau_2\lambda_A\psi)+
({\bar \psi}\tau_2\lambda_A\psi^C)({\bar \psi}^C\tau_2\lambda_A\psi)\right)\ . 
\eeq
Here, $\psi^C=C{\bar \psi}^T$ with $C=i\gamma^2\gamma^0$ being the charge conjugation operator. 
Also, $\tau_2$ is the second component of the Pauli matrices representing 
the isospin $su(2)$-generator and 
$\lambda_A$ are the antisymmetric Gell-Mann matrices representing the color $su(3)_c$-generator.  
As is well known, the Lagrangian density ${\cal L}_0+{\cal L}_S$ corresponds to the original NJL model. 
We add ${\cal L}_c$ which can be derived by the Fierz transformation from ${\cal L}_S$ and 
represents the quark-pair interaction. 
Also, we introduce ${\cal L}_T$ which represents the tensor-type interaction between quarks.

In this paper, we concentrate on quark matter at high baryon density where the chiral symmetry 
is restored in the density region considered here. 
Thus, the chiral condensate, $\langle {\bar \psi}\psi \rangle$, is assumed to be equal to zero in this 
high density region. 
Under the mean field approximation,  the above Lagrangian density is recast into 
\beq\label{2}
& &{\cal L}^{MF}={\cal L}_0+{\cal L}_T^{MF}+{\cal L}_c^{MF}\ , \nonumber\\
& &\ {\cal L}_T^{MF}=-F({\bar \psi}\Sigma_3\tau_3\psi)-\frac{F^2}{2G}\ , \nonumber\\
& &\qquad F=-G\langle {\bar \psi}\Sigma_3 \tau_3 \psi\rangle\ , \qquad
\Sigma_3=-i\gamma^1\gamma^2=
\left(
\begin{array}{cc}
\sigma_3 & 0 \\
0 & \sigma_3 
\end{array}\right)\ , \nonumber\\
& &\ {\cal L}_c^{MF}=-\frac{1}{2}\sum_{A=2,5,7}(\Delta {\bar \psi}^Ci\gamma_5\tau_2\lambda_A
+h.c.)-\frac{3\Delta^2}{2G_c}\ , \nonumber\\
& &\qquad
\Delta_A=\Delta_A^*=-G_c\langle{\bar \psi}i\gamma_5\tau_2\lambda_A\psi\rangle \ , \qquad
\Delta=\Delta_2=\Delta_5=\Delta_7\ ,
\eeq
where $h.c.$ represents the Hermitian conjugate term of the preceding one. 
Here, we used Dirac representation for the Dirac gamma matrices and $\sigma_3$ 
represents the third component of the $2\times 2$ Pauli spin matrices. 
The symbol $\langle\cdots \rangle$ represents the expectation value with respect to a vacuum state. 
The expectation value $F$ corresponds to the order parameter of the spin alignment which leads to quark 
ferromagnetization. 
The expectation value $\Delta$ corresponds to the quark-pair condensate which means the existence of the color superconducting phase 
if $\Delta\neq 0$.
Here, in order to insure color symmetry, we assume that all quark-pair condensates have the 
same expectation values, $\Delta_2=\Delta_5=\Delta_7$.

The mean field Hamiltonian density with quark chemical potential $\mu$ is easily obtained as 
\beq\label{3}
& &{\cal H}_{MF}-\mu{\cal N}={\cal K}_{0}+{\cal H}_{T}^{MF}+{\cal H}_c^{MF}\ , \nonumber\\
& &\quad {\cal K}_{0}={\bar \psi}(-{\mib \gamma}\cdot{\mib \nabla}-\mu\gamma_0)\psi\ , \nonumber\\
& &\quad {\cal H}_T^{MF}=-{\cal L}_T^{MF}\ , \qquad
{\cal H}_c^{MF}=-{\cal L}_c^{MF}
\eeq
with ${\cal N}=\psi^{\dagger}\psi$. 
In the Dirac representation for the Dirac gamma matrices, the Hamiltonian matrix 
of the spin polarization part, $H_{MF}^{SP}=\int d^3{\mib x}\ ({\cal K}_0+{\cal H}_T^{MF})$, 
is written as 
\beq\label{a1}
h_{MF}^{SP}&=&{\mib p}\cdot{\mib \alpha}+F\tau_3\beta \Sigma_3\nonumber\\
&=&
\left(
\begin{array}{cccc}
F\tau_3 & 0 & p_3 & p_1-ip_2 \\
0 & -F\tau_3 & p_1+ip_2 & p_3 \\
p_3 & p_1-ip_2 & -F\tau_3 & 0 \\
p_1+ip_2 & -p_3 & 0 & F\tau_3 
\end{array}
\right)\ , 
\eeq 
where $\alpha^i=\gamma^0\gamma^i$ and $\beta=\gamma^0$. 
For good helicity states, this Hamiltonian matrix is diagonalized with $F=0$. 
For simplicity, we rotate around $p_3$ axis and we set $p_2=0$ without loss of generality. 
In this case, we derive $\kappa=U^{-1}h_{MF}^{SP}U$ as follows: 
\beq\label{a2}
U&=&
\frac{1}{2\sqrt{p}}
\left(
\begin{array}{cccc}
\sqrt{p+p_3} & \sqrt{p-p_3} & -\sqrt{p+p_3} & -\sqrt{p-p_3} \\
\frac{p_1}{|p_1|}\sqrt{p-p_3} & -\frac{p_1}{|p_1|}\sqrt{p+p_3} & -\frac{p_1}{|p_1|}\sqrt{p-p_3} & \frac{p_1}{|p_1|}\sqrt{p+p_3} \\
\sqrt{p+p_3} & -\sqrt{p-p_3} & \sqrt{p+p_3} & -\sqrt{p-p_3} \\
\frac{p_1}{|p_1|}\sqrt{p-p_3} & \frac{p_1}{|p_1|}\sqrt{p+p_3} & \frac{p_1}{|p_1|}\sqrt{p-p_3} & \frac{p_1}{|p_1|}\sqrt{p+p_3} \\
\end{array}
\right)\ , \nonumber\\
\kappa&=&U^{-1}h_{MF}^{SP}U \nonumber\\
&=&\left(
\begin{array}{cccc}
p & 0 & 0 & 0 \\
0 & p & 0 & 0 \\
0 & 0 & -p & 0 \\
0 & 0 & 0 & -p \\
\end{array}
\right)+
\frac{F\tau_3}{p}
\left(
\begin{array}{cccc}
0  & |p_1| & -p_3  & 0 \\
|p_1| & 0 & 0 & p_3 \\
-p_3 & 0 & 0 & |p_1| \\
0 & p_3 & |p_1| & 0 
\end{array}
\right)\ . 
\eeq
Finally, in the original basis rotated around $p_3$-axis, $|p_1|$ is replaced to $\sqrt{p_1^2+p_2^2}$. 
As for the Hamiltonian matrix of the color superconducting part, $H_c^{MF}=\int d^3{\mib x}{\cal H}_c^{MF}$, 
the result was already obtained in Ref.\citen{Kitazawa} based on \citen{2}. 
As a result, in the basis of good helicity states, the relevant combination of the mean field Hamiltonian 
$H_{MF}=\int d^3{\mib x}{\cal H}_{MF}$ and the quark number $N=\int d^3{\mib x}{\cal N}$ is given by 
\beq\label{4}
H_{MF}-\mu N&=&
\sum_{{\mib p}\eta\tau\alpha}\left[(p-\mu)c^{\dagger}_{{\mib p}\eta\tau\alpha}c_{{\mib p}\eta\tau\alpha}
-(p+\mu){\tilde c}^{\dagger}_{{\mib p}\eta\tau\alpha}{\tilde c}_{{\mib p}\eta\tau\alpha}\right]\nonumber\\
& &+F\sum_{{\mib p}\eta\tau\alpha}\phi_\tau\biggl[
\frac{\sqrt{p_1^2+p_2^2}}{p}\left(c^{\dagger}_{{\mib p}\eta\tau\alpha}c_{{\mib p}-\eta\tau\alpha}
+{\tilde c}^{\dagger}_{{\mib p}\eta\tau\alpha}{\tilde c}_{{\mib p}-\eta\tau\alpha}\right)\nonumber\\
& &\qquad\qquad\qquad
-\eta\frac{p_3}{p}\left(c^{\dagger}_{{\mib p}\eta\tau\alpha}{\tilde c}_{{\mib p}\eta\tau\alpha}
+{\tilde c}^{\dagger}_{{\mib p}\eta\tau\alpha}c_{{\mib p}\eta\tau\alpha}\right)\biggl]\nonumber\\
& &+\frac{\Delta}{2}\sum_{{\mib p}\eta\alpha\alpha'\alpha''\tau\tau'}
(c^{\dagger}_{{\mib p}\eta\tau\alpha}c^{\dagger}_{-{\mib p}\eta\tau'\alpha'}
+{\tilde c}^{\dagger}_{{\mib p}\eta\tau\alpha}{\tilde c}^{\dagger}_{-{\mib p}\eta\tau'\alpha'}\nonumber\\
& &\qquad\qquad\qquad
+c_{-{\mib p}\eta\tau'\alpha'}c_{{\mib p}\eta\tau\alpha}
+{\tilde c}_{-{\mib p}\eta\tau'\alpha'}{\tilde c}_{{\mib p}\eta\tau\alpha})\phi_{\tau}\epsilon_{\alpha\alpha'\alpha''}\epsilon_{\tau\tau'}
\nonumber\\
& &+V\cdot\frac{F^2}{2G}+V\cdot\frac{3\Delta^2}{2G_c}\ , 
\eeq
where $V$ represents the volume in the box normalization.\footnote{
We hope that no confusion may occur although notations $V_{cs}$ and $V({\mib p})$ appear later which represent 
the quark-pair interaction in (\ref{5}) and a matrix in (\ref{9}) diagonalizing a certain part of the Hamiltonian matrix, respectively. 
}
Here, $c^{\dagger}_{{\mib p}\eta\tau\alpha}$ and ${\tilde c}^{\dagger}_{{\mib p}\eta\tau\alpha}$ represent 
the quark and antiquark creation operators with momentum ${\mib p}$, helicity $\eta=\pm$, isospin index $\tau=\pm$ and color $\alpha$. 
Further, $\phi_{\tau}=1$ for $\tau=1$ (up quark) and $\phi_{\tau}=-1$ for $\tau=-1$ (down quark). 
Also, $\epsilon_{\tau\tau'}$ and $\epsilon_{\alpha\alpha'\alpha''}$ represent the complete antisymmetric tensor for 
the isospin and color indices. 
We define $p=\sqrt{p_1^2+p_2^2+p_3^2}$, that is, the magnitude of momentum.

For later convenience, we rewrite the above Hamiltonian as 
\beq\label{5}
& &H_{MF}-\mu N=H_{MF}^{SP}-\mu N+V_{cs}+V\cdot\frac{F^2}{2G}+V\cdot\frac{3\Delta^2}{2G_c}
% \nonumber\\
\eeq
with
\beq\label{5add}
& &H_{MF}^{SP}-\mu N=\sum_{{\mib p}\tau\tau'\alpha\alpha'\alpha''}
\left(
\begin{array}{c}
c^{\dagger}_{{\mib p}+\tau\alpha}\\
c^{\dagger}_{{\mib p}-\tau\alpha}\\
{\tilde c}^{\dagger}_{{\mib p}+\tau\alpha}\\
{\tilde c}^{\dagger}_{{\mib p}-\tau\alpha}
\end{array}
\right)^{T}
(\kappa-\mu\cdot 1)\delta_{\tau\tau'}\delta_{\alpha\alpha'}
\left(
\begin{array}{c}
c_{{\mib p}+\tau'\alpha'} \\ 
c_{{\mib p}-\tau'\alpha'} \\
{\tilde c}_{{\mib p}+\tau'\alpha'} \\
{\tilde c}_{{\mib p}-\tau'\alpha'}
\end{array}
\right)\ , \nonumber\\
& &\kappa=
\left(
\begin{array}{cccc}
p & 0 & 0 & 0 \\
0 & p & 0 & 0 \\
0 & 0 & -p & 0 \\
0 & 0 & 0 & -p \\
\end{array}
\right) 
+
\frac{F\tau_3}{p}
\left(
\begin{array}{cccc}
0  & \sqrt{p_1^2+p_2^2} & -p_3  & 0 \\
\sqrt{p_1^2+p_2^2} & 0 & 0 & p_3 \\
-p_3 & 0 & 0 & \sqrt{p_1^2+p_2^2} \\
0 & p_3 & \sqrt{p_1^2+p_2^2} & 0 
\end{array}
\right)\ , \nonumber\\
& &V_{cs}=\frac{\Delta}{2}\sum_{{\mib p}\eta\alpha\alpha'\alpha''\tau\tau'}
(c^{\dagger}_{{\mib p}\eta\tau\alpha}c^{\dagger}_{-{\mib p}\eta\tau'\alpha'}
+{\tilde c}^{\dagger}_{{\mib p}\eta\tau\alpha}{\tilde c}^{\dagger}_{-{\mib p}\eta\tau'\alpha'}
+c_{-{\mib p}\eta\tau'\alpha'}c_{{\mib p}\eta\tau\alpha}
+{\tilde c}_{-{\mib p}\eta\tau'\alpha'}{\tilde c}_{{\mib p}\eta\tau\alpha})
\nonumber\\
& &\qquad\qquad\qquad\qquad
\times \phi_{\tau}\epsilon_{\alpha\alpha'\alpha''}\epsilon_{\tau\tau'}\ . 
\eeq
As for $V_{cs}$, we can easily sum up with respect to $\tau$ and $\tau'$ which leads to 
\beq\label{6}
V_{cs}=\Delta\sum_{{\mib p}\eta\alpha\alpha'\alpha''}
\epsilon_{\alpha\alpha'\alpha''}(c^{\dagger}_{{\mib p}\eta +\alpha}c^{\dagger}_{-{\mib p}\eta -\alpha'}
+{\tilde c}^{\dagger}_{{\mib p}\eta +\alpha}{\tilde c}^{\dagger}_{-{\mib p}\eta -\alpha'}
%\nonumber\\
%& &\qquad\qquad\qquad\qquad
+c_{-{\mib p}\eta -\alpha'}c_{{\mib p}\eta +\alpha}
+{\tilde c}_{-{\mib p}\eta -\alpha'}{\tilde c}_{{\mib p}\eta +\alpha})\ .\nonumber\\
& & 
\eeq

\subsection{Expressions of the Hamiltonian under other basis sets}

If the quark-pair condensate $\Delta$ is equal to zero, the operator given by equation (\ref{5}) becomes 
$H_{MF}^{SP}-\mu N +VF^2/(2G)$. 
Since the quark number $N$ is already diagonal, let us diagonalize the Hamiltonian matrix $\kappa$. 
First, we set up $\tau_3=1$ because the contribution of $\tau_3=-1$ is the same as the contribution of $\tau_3=1$ as was seen 
in (\ref{6}), which results in a factor 2. 
Then, the Hamiltonian matrix $\kappa$ can be simply expressed as 
\beq\label{7}
\kappa=\left(
\begin{array}{cccc}
q & e & -g & 0 \\
e & q & 0 & g \\
-g & 0 & -q & e \\
0 & g & e & -q
\end{array}
\right)\ ,
\eeq  
where $q=p$, $e=F\sqrt{p_1^2+p_2^2}/p$ and $g=Fp_3/p$. 
The eigenvalues of $\kappa$ are easily obtained as 
\beq\label{8}
\pm\varepsilon_{\mib p}^{(\pm)}
=\pm\sqrt{g^2+(e\pm q)^2}=\pm\sqrt{p_3^2+\left(F\pm\sqrt{p_1^2+p_2^2}\right)^2}\ . 
\eeq

By introducing new fermion operators 
$(a_{{\mib p}\eta \tau \alpha}, a^{\dagger}_{{\mib p}\eta \tau \alpha},
{\tilde a}_{{\mib p}\eta \tau \alpha}, {\tilde a}^{\dagger}_{{\mib p}\eta \tau \alpha})$ by 
\beq\label{16}
\left(
\begin{array}{c}
a_{{\mib p}+\tau\alpha}\\
a_{{\mib p}-\tau\alpha}\\
{\tilde a}_{{\mib p}+\tau\alpha}\\
{\tilde a}_{{\mib p}-\tau\alpha}
\end{array}
\right)
=W^{\dagger}V^{\dagger}({\mib p})
\left(
\begin{array}{c}
c_{{\mib p}+\tau\alpha}\\
c_{{\mib p}-\tau\alpha}\\
{\tilde c}_{{\mib p}+\tau\alpha}\\
{\tilde c}_{{\mib p}-\tau\alpha}
\end{array}
\right) \ , 
\eeq
where the operators $V({\mib p})$ and $W$ are given in the Appendix A, 
the mean field Hamiltonian in which both quark spin polarization and quark-pair condensate 
are simultaneously considered can be expressed as 
\beq\label{17}
H_{MF}-\mu N
&=&
\sum_{{\mib p}\eta\tau\alpha}
\left[
(\varepsilon_{\mib p}^{(\eta)}-\mu)a^{\dagger}_{{\mib p}\eta\tau\alpha}a_{{\mib p}\eta\tau\alpha}
-(\varepsilon_{\mib p}^{(\eta)}+\mu){\tilde a}^{\dagger}_{{\mib p}\eta\tau\alpha}
{\tilde a}_{{\mib p}\eta\tau\alpha}\right]
\nonumber\\
& &+\frac{\Delta}{2}
\sum_{{\mib p}\eta\tau\tau'\alpha\alpha'\alpha''}
\biggl[
a^{\dagger}_{{\mib p}\eta\tau\alpha'}a^{\dagger}_{-{\mib p}\eta\tau'\alpha''}
-
{\tilde a}^{\dagger}_{{\mib p}\eta\tau\alpha'}{\tilde a}^{\dagger}_{-{\mib p}\eta\tau\alpha''}
+h.c.
\biggl]\epsilon_{\alpha\alpha'\alpha''}\epsilon_{\tau\tau'}\phi_{\tau} 
\nonumber\\
& &+V\cdot\frac{F^2}{2G}+V\cdot\frac{3\Delta^2}{2G_c}\ . 
\eeq  
Here, the summation with respect to isospin indices $\tau$ and $\tau'$ is explicitly rewritten. 
The above mean field Hamiltonian is the starting point for discussing the quark spin polarized phase and 
color superconducting phase.

\section{BCS state and thermodynamic potential}

We introduce the BCS state following Ref.\citen{3}:
\beq\label{18}
& &\ket{\Psi}=e^S\ket{\Psi_0}\ , \qquad
\ket{\Psi_0}=\prod_{
{\mib p}\eta\tau\alpha(\varepsilon_{\mib p}^{(\eta)}<\mu)
}
a^{\dagger}_{{\mib p}\eta\tau\alpha}\ket{0}\ , \nonumber\\
& &S=\sum_{{\mib p}\eta (\varepsilon_{\mib p}^{(\eta)}>\mu)}
\frac{K_{{\mib p}\eta}}{2}
\sum_{\alpha\alpha'\alpha''\tau\tau'}
a^{\dagger}_{{\mib p}\eta\tau\alpha}a^{\dagger}_{-{\mib p}\eta\tau'\alpha'}
\epsilon_{\alpha\alpha'\alpha''}\epsilon_{\tau\tau'}\phi_{\tau}\nonumber\\
& &\qquad
+\sum_{{\mib p}\eta (\varepsilon_{\mib p}^{(\eta)}\leq \mu)}
\frac{{\wtilde K}_{{\mib p}\eta}}{2}
\sum_{\alpha\alpha'\alpha''\tau\tau'}
a_{{\mib p}\eta\tau\alpha}a_{-{\mib p}\eta\tau'\alpha'}
\epsilon_{\alpha\alpha'\alpha''}\epsilon_{\tau\tau'}\phi_{\tau}\ ,
\eeq
where $\ket{0}$ is the vacuum state with respect to $a_{{\mib p}\eta\tau\alpha}$ and 
$K_{{\mib p}\eta}=K_{-{\mib p}\eta}$ and ${\wtilde K}_{{\mib p}\eta}={\wtilde K}_{-{\mib p}\eta}$ are satisfied. 
Here, the contribution of quark-pairing with negative energy is not considered 
in quark matter.  
Then, the state $\ket{\Psi}$ is a vacuum state with respect to new operators 
$d_{{\mib p}\eta\tau\alpha}$: 
\beq\label{19}
& &d_{{\mib p}\eta\tau\alpha}=
\left\{
\begin{array}{ll}
a_{{\mib p}\eta\tau\alpha}-K_{{\mib p}\eta}(a^{\dagger}_{-{\mib p}\eta\tau'\beta}
-a^{\dagger}_{-{\mib p}\eta\tau'\gamma}) & \qquad{\rm for}\quad \varepsilon_{\mib p}^{(\eta)}>\mu 
\\
a^{\dagger}_{{\mib p}\eta\tau\alpha}+{\wtilde K}_{{\mib p}\eta}(a_{-{\mib p}\eta\tau'\beta}
-a_{-{\mib p}\eta\tau'\gamma}) & \qquad{\rm for}\quad \varepsilon_{\mib p}^{(\eta)}\leq\mu 
\end{array}
\right.\ , 
\eeq
where $(\alpha, \beta, \gamma)$ is a cyclic permutation for color indices and $\tau'=-\tau$. 
For the above operators, the state $\ket{\Psi}$ satisfies the following relation:
\beq\label{20}
d_{{\mib p}\eta\tau\alpha}\ket{\Psi}=0\ .
\eeq
The state $\ket{\Psi}$ is identical with the BCS state.

In order to calculate the expectation value of the mean field Hamiltonian (\ref{17}), 
it is necessary to obtain the expectation values of 
$a^{\dagger}_{{\mib p}\eta\tau\alpha}
a_{{\mib p}\eta\tau\alpha}$, 
$a^{\dagger}_{{\mib p}\eta\tau\alpha}a^{\dagger}_{-{\mib p}\eta\tau'\alpha'}$ and so on 
with respect to the BCS state $\ket{\Psi}$. 
The detailed calculations are given in Appendix B. 
Thus, by summing up with respect to color and isospin indices, 
we obtain the expectation value of the mean field Hamiltonian for the BCS state as 
\beq\label{26}
\bra{\Psi}
H_{MF}-\mu N\ket{\Psi}
&=&
3\sum_{{\mib p}\eta\tau(\varepsilon_{\mib p}^{(\eta)}>\mu)}
\left[
(\varepsilon_{\mib p}^{(\eta)}-\mu)\frac{2K_{{\mib p}\eta}^2}{1+3K_{{\mib p}\eta}^2}
+2\Delta\frac{K_{{\mib p}\eta}}{1+3K_{{\mib p}\eta}^2}\right]
\nonumber\\
& &+
3\sum_{{\mib p}\eta\tau(\varepsilon_{\mib p}^{(\eta)}\leq\mu)}
\left[
(\varepsilon_{\mib p}^{(\eta)}-\mu)
\left(1-\frac{2{\wtilde K}_{{\mib p}\eta}^2}{1+3{\wtilde K}_{{\mib p}\eta}^2}
\right)
+2\Delta\frac{{\wtilde K}_{{\mib p}\eta}}{1+3{\wtilde K}_{{\mib p}\eta}^2}\right]
\nonumber\\
& &+V\cdot\frac{F^2}{2G}+V\cdot\frac{3\Delta^2}{2G_c}\ . 
\eeq
In order to determine the BCS state, namely, to determine the variational variables 
$K_{{\mib p}\eta}$ and ${\wtilde K}_{{\mib p}\eta}$, we introduce new variational variables 
$\theta_{p\eta}$ and ${\tilde \theta}_{p\eta}$ instead of 
$K_{{\mib p}\eta}$ and ${\wtilde K}_{{\mib p}\eta}$:
\beq\label{27}
& &\sin\theta_{p\eta}=\frac{\sqrt{3}K_{{\mib p}\eta}}{\sqrt{1+3K_{{\mib p}\eta}^2}}\ , \qquad
\cos\theta_{p\eta}=\frac{1}{\sqrt{1+3K_{{\mib p}\eta}^2}}\ , \nonumber\\
& &\sin{\tilde \theta}_{p\eta}
=\frac{\sqrt{3}{\wtilde K}_{{\mib p}\eta}}{\sqrt{1+3{\wtilde K}_{{\mib p}\eta}^2}}\ , \qquad
\cos{\tilde \theta}_{p\eta}=\frac{1}{\sqrt{1+3{\wtilde K}_{{\mib p}\eta}^2}}\ . 
\eeq
Thus, the Hamiltonian and color superconducting gap $\Delta$ can be expressed in terms 
of $\theta_{p\eta}$ and ${\tilde \theta}_{p\eta}$ as 
\beq
\bra{\Psi}
H_{MF}-\mu N\ket{\Psi}
&=&2\sum_{{\mib p}\eta (\varepsilon_{\mib p}^{(\eta)}>\mu)}
\left[
2(\varepsilon_{\mib p}^{(\eta)}-\mu)\sin^2\theta_{p\eta}
+2\sqrt{3}\Delta\sin\theta_{p\eta}\cos\theta_{p\eta}\right]
\nonumber\\
& &+
2\sum_{{\mib p}\eta (\varepsilon_{\mib p}^{(\eta)}\leq\mu)}
\left[
(\varepsilon_{\mib p}^{(\eta)}-\mu)(3-2\sin^2{\tilde \theta}_{p\eta})
+2\sqrt{3}\Delta\sin{\tilde \theta}_{p\eta}\cos{\tilde \theta}_{p\eta}\right]
\nonumber\\
& &+V\cdot\frac{F^2}{2G}+V\cdot\frac{3\Delta^2}{2G_c}\ , 
\label{28}\\
\Delta&=&\Delta_2=\Delta_5=\Delta_7
=-G_c\cdot\frac{1}{V}\sum_{{\mib p}\eta\beta\gamma\tau}
\bra{\Psi}a^{\dagger}_{{\mib p}\eta\tau\beta}a^{\dagger}_{-{\mib p}\eta -\tau \gamma}\ket{\Psi}
\nonumber\\
&=&-2G_c\left(
\frac{1}{V}\!\!\sum_{{\mib p}\eta\tau(\varepsilon_{\mib p}^{(\eta)}>\mu)}
\frac{K_{{\mib p}\eta}}{1+3K_{{\mib p}\eta}^2}+
\frac{1}{V}\!\!\sum_{{\mib p}\eta\tau(\varepsilon_{\mib p}^{(\eta)}\leq \mu)}
\frac{{\wtilde K}_{{\mib p}\eta}}{1+3{{\wtilde K}}_{{\mib p}\eta}^2}\right)\nonumber\\
&=&-\frac{4G_c}{\sqrt{3}}\left(\!\!
\frac{1}{V}\!\!\!\sum_{{\mib p}\eta(\varepsilon_{\mib p}^{(\eta)}>\mu)}
\!\!\sin\theta_{p\eta}\cos\theta_{p\eta}+
\frac{1}{V}\!\!\!\sum_{{\mib p}\eta(\varepsilon_{\mib p}^{(\eta)}\leq \mu)}
\!\!\sin{\tilde \theta}_{p\eta}\cos{\tilde \theta}_{p\eta}\!\!\right) .\qquad
\label{29}
\eeq
Here, in the first line in (\ref{28}) and the third line in (\ref{29}), the 
isospin indices are summed up and an extra factor 2 appears with respect to the corresponding expression in \citen{2}. 
Also, in the second line in (\ref{29}), color 
indices, $\beta$ and $\gamma$, are summed up and an extra factor 2 also appears.

Next, we impose the minimization condition for $\bra{\Psi} H_{MF}-\mu N \ket{\Psi}$ with 
respect to $\Delta$, $\theta_{p\eta}$ and ${\tilde \theta}_{p\eta}$. 
First, we impose the minimization condition with respect to $\Delta$: 
\beq\label{30}
\frac{\partial}{\partial \Delta}\bra{\Psi}H_{MF}-\mu N\ket{\Psi}
=0\ .
\eeq
This minimization condition leads to the gap equation in (\ref{29}) exactly. 
Secondly, we minimize $\bra{\Psi} H_{MF}-\mu N\ket{\Psi}$ with respect to 
$\theta_{p\eta}$ and ${\tilde \theta}_{p\eta}$:
\beq\label{31}
\frac{\partial}{\partial \theta_{p\eta}}\bra{\Psi}H_{MF}-\mu N\ket{\Psi}
=0\ , \qquad
\frac{\partial}{\partial {\tilde \theta}_{p\eta}}\bra{\Psi}H_{MF}-\mu N\ket{\Psi}
=0 \ , 
\eeq
which lead to the following equations:
\begin{equation}\label{32}
\begin{split}
(\varepsilon_{\mib p}^{(\eta)}-\mu)\sin 2\theta_{p\eta}+\sqrt{3}\Delta \cos 2\theta_{p\eta}& =0 \ ,
\\
-(\varepsilon_{\mib p}^{(\eta)}-\mu)\sin 2{\tilde \theta}_{p\eta}
+\sqrt{3}\Delta \cos 2{\tilde \theta}_{p\eta}& =0 \ .
\end{split}
\end{equation}
Thus, we obtain 
$\tan 2\theta_{p\eta}=-\sqrt{3}\Delta/(\varepsilon_{\mib p}^{(\eta)}-\mu)$ and 
$\tan 2{\tilde \theta}_{p\eta}=\sqrt{3}\Delta/(\varepsilon_{\mib p}^{(\eta)}-\mu)$ which gives 
\b\label{33}
\bs
& \sin\theta_{p\eta}^2=\frac{1}{2}\left[
1-\frac{\varepsilon_{\mib p}^{(\eta)}-\mu}{\sqrt{
(\varepsilon_{\mib p}^{(\eta)}-\mu)^2+3\Delta^2}}\right]\ , \\
& \sin{\tilde \theta}_{p\eta}^2=\frac{1}{2}\left[
1+\frac{\varepsilon_{\mib p}^{(\eta)}-\mu}{\sqrt{
(\varepsilon_{\mib p}^{(\eta)}-\mu)^2+3\Delta^2}}\right]\ . 
\end{split}
\e
Since the variational parameters $\theta_{p\eta}$ and ${\tilde \theta}_{p\eta}$ are 
determined completely, namely, $K_{{\mib p}\eta}$ and ${\wtilde K}_{{\mib p}\eta}$ are determined, 
the BCS state is obtained. 

Thus, we can derive the thermodynamic potential $\Phi(\Delta,F,\mu)$ at zero temperature from (\ref{28}) 
with (\ref{33}) as 
\beq\label{34}
\Phi(\Delta,F,\mu)
&=&\frac{1}{V}\bra{\Phi}H_{MF}-\mu N\ket{\Phi}\nonumber\\
&=&
2\cdot\frac{1}{V}\sum_{{\mib p}\eta (\varepsilon_{\mib p}^{(\eta)}\leq \mu)}
\left[2(\varepsilon_{\mib p}^{(\eta)}-\mu)-\sqrt{(\varepsilon_{\mib p}^{(\eta)}-\mu)^2+3\Delta^2}
\right]\nonumber\\
& &
+2\cdot\frac{1}{V}\sum_{{\mib p}\eta (\varepsilon_{\mib p}^{(\eta)}> \mu)}^{\Lambda}
\left[(\varepsilon_{\mib p}^{(\eta)}-\mu)-\sqrt{(\varepsilon_{\mib p}^{(\eta)}-\mu)^2+3\Delta^2}
\right]\nonumber\\
%& &
%+2\sum_{{\mib p}\ (\varepsilon_{\mib p}^{(-)}\leq \mu)}
%\left[2(\varepsilon_{\mib p}^{(-)}-\mu)-\sqrt{(\varepsilon_{\mib p}^{(-)}-\mu)^2+3\Delta^3}
%\right]\nonumber\\
%& &
%+2\sum_{{\mib p}\ (\varepsilon_{\mib p}^{(-)}> \mu)}^{\Lambda}
%\left[(\varepsilon_{\mib p}^{(-)}-\mu)-\sqrt{(\varepsilon_{\mib p}^{(-)}-\mu)^2+3\Delta^3}
%\right]\nonumber\\
& &+\frac{F^2}{2G}+\frac{3\Delta^2}{2G_c}\ , 
\eeq
where $\varepsilon_{\mib p}^{(\pm)}=
\sqrt{p_3^2+\left(F\pm\sqrt{p_1^2+p_2^2}\right)^2}$. 
Here, we explicitly introduce and write a three momentum cutoff parameter $\Lambda$. 
The above derivation is nothing but that of the usual method in the BCS theory. 
Therefore, it is possible to derive the same result using the equations of motion 
by eliminating so-called dangerous terms in the algebraic method \cite{3}.
It may be interesting to observe that Eq.(\ref{34}) means, physically, that only two ``effective colors" 
participate in the pairing process, the third ``color" remaining inert. 
The contribution of the active ``colors" to the 
energy is $\left(2(\varepsilon_{\mib p}^{(\eta)}-\mu)-2\sqrt{(\varepsilon_{\mib p}^{(\eta)}-\mu)^2+3\Delta^2}\right)/2$, 
both for 
$\varepsilon_{\mib p}^{(\eta)} < \mu$ and for $\varepsilon_{\mib p}^{(\eta)} >\mu$, while the 
contribution of the inert ``color" is simply $\varepsilon_{\mib p}^{(\eta)}-\mu$, but only for 
$\varepsilon_{\mib p}^{(\eta)} < \mu$.
The gap equation in (\ref{29}), that is $\partial \Phi(\Delta, F, \mu)/\partial \Delta=0$, 
is obtained as
\beq\label{35}
\Delta\left[
2\cdot\frac{1}{V}\sum_{{\mib p}\eta=\pm}^\Lambda \frac{1}{\sqrt{(\varepsilon_{\mib p}^{(\eta)}-\mu)^2+3\Delta^2}}
-\frac{1}{G_c}\right]=0\ . 
\eeq

\section{Numerical results and discussions}

In this section, first, we investigate the possible phases with $F=0$ and $\Delta\neq 0$ or 
with $F\neq 0$ and $\Delta=0$ at high density, namely, two-flavor color superconducting phase or 
spin polarized phase, separately. 
First, as an extreme situation in the quark spin polarized phase with $\Delta=0$, the thermodynamic potential 
reduces to  
\beq\label{36}
\Phi(\Delta=0, F,\mu)
&=&6\cdot\frac{1}{V}\sum_{{\mib p}\ (\varepsilon_{\mib p}^{(+)}\leq \mu)}
(\varepsilon_{\mib p}^{(+)}-\mu)
+6\cdot\frac{1}{V}\sum_{{\mib p}\ (\varepsilon_{\mib p}^{(-)}\leq \mu)}
(\varepsilon_{\mib p}^{(-)}-\mu)+\frac{F^2}{2G}\ , \quad
\eeq
where it should be noted that 
$\sqrt{(\varepsilon_{\mib p}^{(\eta)}-\mu)^2}=
\varepsilon_{\mib p}^{(\eta)}-\mu$ for $\varepsilon_{\mib p}^{(\eta)}>\mu$ or 
$-(\varepsilon_{\mib p}^{(\eta)}-\mu)$ for $\varepsilon_{\mib p}^{(\eta)}<\mu$, 
respectively, when (\ref{36}) is derived from (\ref{34}).  
Of course, the sum over momentum is replaced by the momentum-integration as 
\beq\label{37}
\frac{1}{V}\sum_{\mib p}\longrightarrow \int \frac{d^3{\mib p}}{(2\pi)^3}\ . 
\eeq 
Then, the thermodynamic potential derived here is identical with the one which has been previously obtained by the present authors, 
that is, Eq.(3.1) in Ref. \citen{1}.\footnote{
In Eq.(3.1) in Ref.\cite{1}, $F$-integration has to be carried out. 
}
Thus, the thermodynamic potential with $\Delta=0$ has already been given in \citen{1} in the analytical form as 
\beq\label{38}
\Phi(\Delta=0, F, \mu)
=
\left\{
\begin{array}{ll}
\displaystyle \frac{F^2}{2G}-\frac{1}{\pi^2}
\biggl[
\frac{\sqrt{\mu^2-F^2}}{4}(3F^2\mu+2\mu^3) & 
\displaystyle
\!\!\!\!\!\!
+F\mu^3\arctan\frac{F}{\sqrt{\mu^2-F^2}} \\
 & \\
\displaystyle 
\qquad\quad\qquad
-\frac{F^4}{4}\ln \frac{\mu+\sqrt{\mu^2-F^2}}{F}\biggl] & 
\qquad\qquad {\rm for}\quad F<\mu \\
\displaystyle \frac{F^2}{2G}-\frac{1}{2\pi}\mu^3 F & \qquad\qquad {\rm for}\quad F>\mu
\end{array}\right. \ . 
\eeq
If the minimum of the thermodynamical potential exists in the range of $F<\mu$, the 
gap equation for $F$ is derived from $\partial\Phi(\Delta=0, F,\mu)/\partial F=0$ 
which leads to 
\beq\label{39}
F=\frac{G}{\pi^2}
\biggl[
2F\mu\sqrt{\mu^2-F^2}+\mu^3\arctan\frac{F}{\sqrt{\mu^2-F^2}}
-F^3\ln\frac{\mu+\sqrt{\mu^2-F^2}}{F}\biggl]\ . 
\eeq

In the other case, namely $F=0$ case, the two-flavor color superconducting phase 
may be realized with $\Delta\neq 0$. 
When $F=0$, the quasiparticle energy $\varepsilon_{\mib p}^{(\eta)}=p$ is obtained for 
$\eta=\pm$. 
The thermodynamic potential (\ref{34}) is then evaluated as 
\beq\label{40}
\Phi(\Delta,F=0,\mu)
&=&4\cdot\frac{1}{V}\sum_{p<\mu}\left(2(p-\mu)-\sqrt{(p-\mu)^2+3\Delta^2}\right)
\nonumber\\
& &
+4\cdot\frac{1}{V}\sum_{p>\mu}^\Lambda \left((p-\mu)-\sqrt{(p-\mu)^2+3\Delta^2}\right)
+\frac{3\Delta^2}{2G_c}\ . 
\eeq
After replacement of the sum over momentum into momentum-integration in (\ref{37}), 
the analytical form of the thermodynamic potential is obtained with a three-momentum cutoff parameter $\Lambda$:
\beq\label{41}
\Phi(\Delta, F=0,\mu)
&=&
\frac{3\Delta^2}{2G_c}-\frac{\mu^4}{6\pi^2}+\frac{\Lambda^4}{2\pi^2}-\frac{2\mu\Lambda^3}{3\pi^2}
\nonumber\\
& &
-\frac{1}{12\pi^2}
\biggl[
(2\mu^3-39\Delta^2\mu)(\sqrt{\mu^2+3\Delta^2}-\sqrt{(\mu-\Lambda)^2+3\Delta^2})
\nonumber\\
& &\qquad\quad
+(6\Lambda^3+9\Delta^2\Lambda-2\Lambda^2\mu-2\Lambda\mu^2)
\sqrt{(\mu-\Lambda)^2+3\Delta^2}
\nonumber\\
& &\qquad\quad
+3(12\Delta^2\mu^2-9\Delta^4)
\ln\frac{\Lambda-\mu+\sqrt{(\mu-\Lambda)^2+3\Delta^2}}{-\mu+\sqrt{\mu^2+3\Delta^2}}
\biggl]\ . 
\eeq
The gap equation becomes 
\beq\label{42}
& &\frac{1}{\pi^2}
\biggl[
-3\mu\sqrt{\mu^2+3\Delta^2}+(3\mu+\Lambda)\sqrt{(\mu-\Lambda)^2+3\Delta^2}
\nonumber\\
& &\qquad
+(2\mu^2-3\Delta^2)
\ln\frac{\Lambda-\mu+\sqrt{(\mu-\Lambda)^2+3\Delta^2}}{-\mu+\sqrt{\mu^2+3\Delta^2}}
\biggl]=\frac{1}{G_c}\ . 
\eeq

The quark number density $\rho$ is calculated from 
the thermodynamical relation as 
\beq\label{43}
\rho=-\frac{\partial \Phi(\Delta, F,\mu)}{\partial \mu}\ . 
\eeq
In our model Hamiltonian, the model parameters are $G$, $G_c$ and $\Lambda$. 
In the original NJL model, the coupling strength $G_S$ appears where $G_S=5.5$ GeV$^{-2}$ is adopted \cite{HK} 
although this parameter does not appear explicitly in the 
model considered here. 
%The values of these parameters are justified by the consistency of the physical quantities 
%in this model, namely, the pion decay constant and the quark mass are reasonably reproduced.  
Then, following Ref.\citen{Kitazawa}, the strength of the quark-pair interaction, $G_c$, is taken\footnote{
In Ref.\citen{Kitazawa}, $G_C/G_S=0.6$ is adopted where $G_C$ in Ref.\citen{Kitazawa} corresponds to $G_c/2$ here.  
} 
as $(G_c/2)/G_S=0.6$.
Thus, we adopt $G_c=6.6$ GeV$^{-2}$. 
As for the three-momentum cutoff $\Lambda$, a standard value \cite{HK} is adopted here, namely, 
$\Lambda=0.631$ GeV. 
If the tail stretching beyond the Fermi momentum of occupation number should be fully taken into account in the BCS theory, 
a larger value of three-momentum cutoff should be adopted such as $\Lambda=0.8$ GeV \cite{3}. 
However, it will be later seen that the tail of the occupation number is already sufficiently taken into account in the case 
$\Lambda=0.631$ GeV.   
As for $G$ in the strength of the tensor-type interaction between quarks, we put $G=20$ GeV$^{-2}$, 
which was used in our previous paper \cite{1}.  
As was discussed in Ref.\citen{1}, if the effect of the vacuum polarization is taken into account, 
the coupling constant $G$ should be replaced to the renormalized coupling $G_r$ in which 
$1/G_r=1/G-\Lambda^2/\pi^2$. 
Then, $G_r=20$ GeV$^{-2}$ corresponds to the bare coupling $G=11.1$ GeV$^{-2}$ for $\Lambda=0.631$ GeV or 
$G=8.7$ GeV$^{-2}$ for $\Lambda=0.8$ GeV. 
Thus, the strengths of the quark-pairing and tensor-type interactions are comparable. 
The parameter set used here is summarized in Table 1. 

%%%%%%%%%%%%%%%%%%%%%%%%%%%%%%%%%%%%%%%%%%%%%%%%%%%%%%%%%%%%%%%%%%%
\begin{table}[t]
\caption{Parameter set}%%%Table caption goes here
\label{table1}
\centering
\begin{tabular}{|c|c|c|}%%%The number of columns has to be defined here
\hline
$\Lambda$ / GeV & $G$ / GeV$^{-2}$ & $G_c$ / GeV$^{-2}$\\ 
\hline
0.631 & 20.0 & 6.6 \\
%%%% Table body
\hline
%Parameter set II & 0.800 & 20.0 & 6.6 \\
%\hline
\end{tabular}
\end{table}%%%End of the table
%%%%%%%%%%%%%%%%%%%%%%%%%%%%%%%%%%%%%%%%%%%%%%%%%%%%%%%%%%%%%%%%%%%%%%%%%%%%%%%%%%%%%%%%

First, let us estimate the thermodynamic potential numerically in two phases, 
namely, color superconducting phase with $F=0$ and $\Delta \neq 0$ and 
quark spin polarized phase with $F\neq 0$ and $\Delta=0$, separately. 
The state with $(F=0, \Delta=\Delta_0)$ gives a local minimum of the thermodynamic potential, 
where $\Delta_0$ is the solution of gap equation (\ref{35}) with $F=0$ or Eq.(\ref{42}). 
However, the state with $(F=F_0, \Delta=0)$ gives a local minimum or a saddle point 
of the thermodynamic potential corresponding to the value of $\mu$, 
where $F_0$ is the solution of gap equation 
$\partial \Phi(\Delta=0, F,\mu)/\partial F=0$ or Eq.(\ref{39}). 
This is shown in Appendix C.
We compare the pressure in the color superconducting phase $(F=0, \Delta=\Delta_0)$ 
with the one in the quark spin polarized phase $(F=F_0, \Delta=0)$. 
The pressure $p$ is given by 
\beq\label{44}
p=-\Phi(\Delta,F,\mu)\ . 
\eeq
In Fig.1 (a), the pressures for normal (thin curve), two-flavor color superconducting (dash-dotted curve) and 
quark spin polarized (solid curve) phases are shown in the case of $\Lambda=0.631$ GeV as functions with respect to quark chemical potential $\mu$. 
Up to $\mu=\mu_c=0.442$ GeV, where $\Phi(\Delta=\Delta_0,F=0,\mu_c)=\Phi(\Delta=0,F=F_0,\mu_c)$, 
the two-flavor color superconducting (2SC) phase is realized. 
However, above $\mu=\mu_c$, the quark spin polarized phase is favored. 
In Fig. 1 (b), details are depicted around $\mu\approx 0.442 (=\mu_c)$ GeV. 
%The value of chemical potential in the phase transition point is 
%0.442 GeV, which leads to the 2SC phase up to $\mu=0.442(=\mu_c)$ GeV. 
As for the baryon number density, up to $\rho_B=4.73 \rho_0$, where $\rho_0=0.17$ fm${}^{-3}$ is the normal nuclear matter density, 
the color superconducting phase is realized. 
Thus, it is enough to include the effects of the tail of occupation number with the three-momentum cutoff 
$\Lambda=0.631$ GeV. 
In Fig.\ref{fig:occupation}, the occupation number is depicted as a function of the magnitude of 
momentum $p(=|{\mib p}|)$. It seems from Fig.\ref{fig:occupation} that 
the effects of the tail of occupation number are fully taken into account.

%%%%%%%%%%%%%%%%%%%%%%%%%%%%%%%%%%%%%%%%%%%%%%%%%%%%%%%%%%%%%%%%%%%%%%
\begin{figure}[t]
\begin{center}
\includegraphics[height=5.2cm]{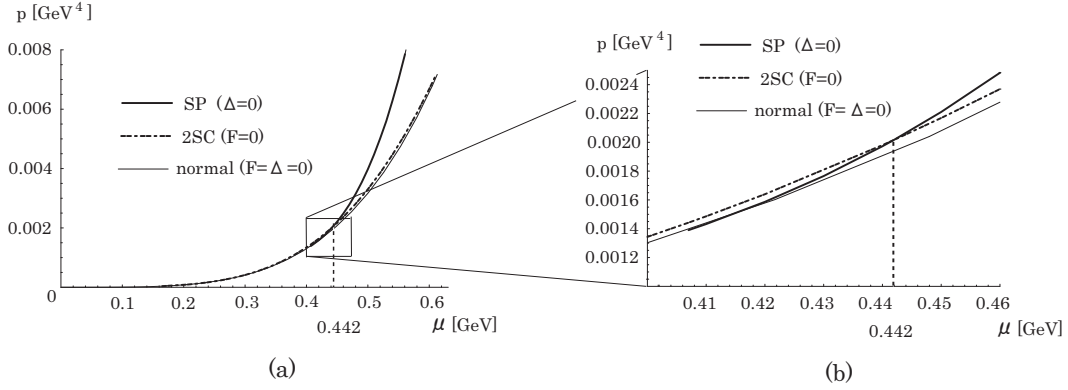}
\caption{(a) The pressures for normal (thin curve), two-flavor color superconducting (2SC) (dash-dotted curve) and 
quark spin polarized (SP) (solid curve) phases are shown as functions with respect to quark chemical potential $\mu$ 
in the case of $\Lambda=0.631$ GeV . 
(b) The details are depicted around $\mu\approx 0.442$ GeV. 
}
\label{fig:1}
\end{center}
\end{figure}
%%%%%%%%%%%%%%%%%%%%%%%%%%%%%%%%%%%%%%%%%%%%%%%%%%%%%%%%%%%%%%%%%%%%%%%%

%%%%%%%%%%%%%%%%%%%%%%%%%%%%%%%%%%%%%%%%%%%%%%%%%%%%%%%%%%%%%%%%%%%%%%
%\begin{figure}[b]
%\begin{center}
%\includegraphics[height=5.0cm]{PvsRho_631.eps}
%\caption{The realized phase is shown by the solid curve in the case of $\Lambda=0.631$ GeV. 
%The vertical axis represents the pressure and the horizontal axis represents the baryon number density divided by the 
%normal nuclear density $\rho_0=0.17$ fm$^{-3}$. 
%}
%\label{fig:2}
%\end{center}
%\end{figure}
%%%%%%%%%%%%%%%%%%%%%%%%%%%%%%%%%%%%%%%%%%%%%%%%%%%%%%%%%%%%%%%%%%%%%%%%

%%%%%%%%%%%%%%%%%%%%%%%%%%%%%%%%%%%%%%%%%%%%%%%%%%%%%%%%%%%%%%%%%%%%%%
\begin{figure}[b]
\begin{center}
\includegraphics[height=4.5cm]{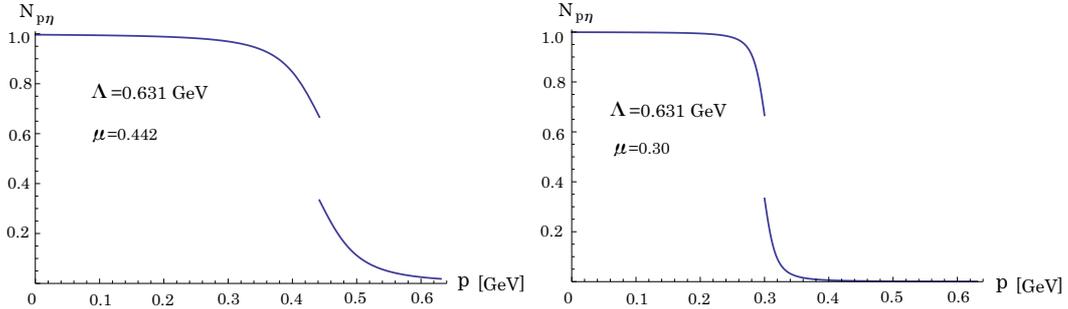}
\caption{The occupation number is depicted as a function of $|{\mib p}|$ in the cases
$\mu=0.442$ GeV and $\mu=0.30$ GeV. 
}
\label{fig:occupation}
\end{center}
\end{figure}
%%%%%%%%%%%%%%%%%%%%%%%%%%%%%%%%%%%%%%%%%%%%%%%%%%%%%%%%%%%%%%%%%%%%%%%%

%%%%%%%%%%%%%%%%%%%%%%%%%%%%%%%%%%%%%%%%%%%%%%%%%%%%%%%%%%%%%%%%%%%%%%
\begin{figure}[t]
\begin{center}
\includegraphics[height=16.7cm]{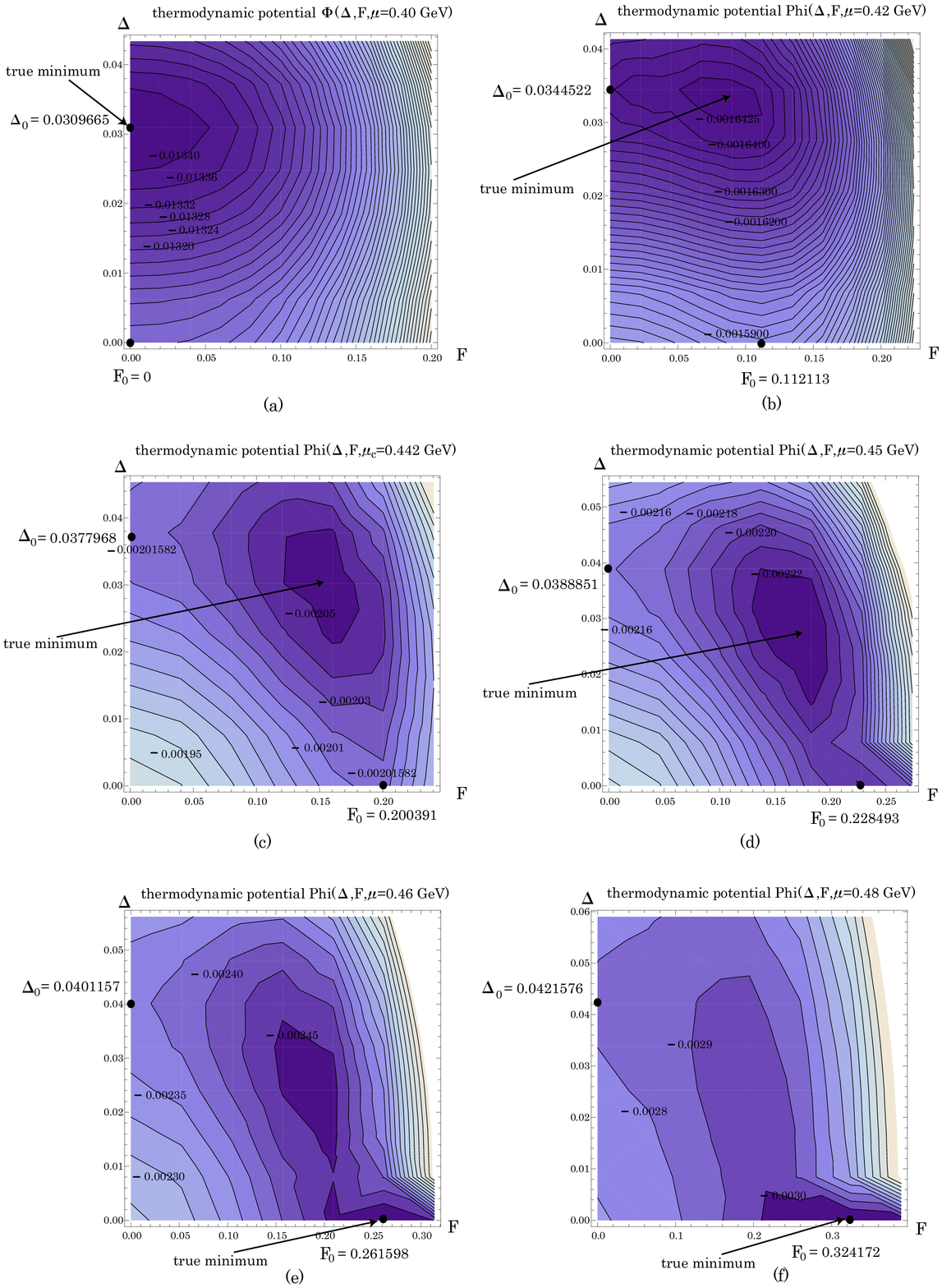}
\caption{The contour map of the thermodynamic potential $\Phi(\Delta,F,\mu)$ is depicted in the cases 
(a) $\mu=0.40$ GeV, (b) $\mu=0.42$ GeV, (c) $\mu=0.442$ GeV, (d) $\mu=0.45$ GeV, (e) $\mu=0.46$ GeV 
and (f) $\mu=0.48$ GeV. 
The horizontal and vertical axis represent 
$F$ and $\Delta$, respectively. 
%The true minimum is located at $\Delta=\Delta_0$ and $F=0$. 
}
\label{fig:contour}
\end{center}
\end{figure}
%%%%%%%%%%%%%%%%%%%%%%%%%%%%%%%%%%%%%%%%%%%%%%%%%%%%%%%%%%%%%%%%%%%%%%%%

Secondly, let us consider the gap $\Delta$ and $F$ simultaneously. 
Figure \ref{fig:contour} shows contour maps of the thermodynamic potential 
$\Phi(\Delta,F,\mu)$, where the horizontal and vertical axis represent 
$F$ and $\Delta$, respectively, 
in the cases (a) $\mu=0.40$ GeV, (b) $\mu=0.42$ GeV, (c) $\mu=\mu_c=0.442$ GeV, (d) $\mu=0.45$ GeV, (e) $\mu=0.46$ GeV 
and (f) $\mu=0.48$ GeV. 
In the case (a), the gap equations gives $F=0$ and $\Delta=\Delta_0$. 
The point $(F=0, \Delta=\Delta_0)$ is a true minimum of the thermodynamic potential. 
Then, up to $\mu\approx 0.40$ GeV, the color superconducting phase is realized. 
In the region with $\mu \gsim 0.407$ GeV which corresponds to the baryon number density $\rho$ divided by the normal 
nuclear matter density $\rho_0$ being $\rho/\rho_0 =3.71$, 
the gap equation $\partial \Phi(\Delta=0,F,\mu)/\partial F$ has 
a nontrivial solution with non-zero $F$ value. 
Then, the true minimum of the thermodynamic potential moves to the point $(F\neq 0, \Delta\approx \Delta_0)$ from the point 
$(F=0,\Delta=\Delta_0)$ as is seen in Fig.\ref{fig:contour}(b). 
In Fig.\ref{fig:contour} (c) with $\mu=\mu_{c}=0.442$ GeV, 
in which $\Phi(\Delta=\Delta_0,F=0,\mu=\mu_{c})=\Phi(\Delta=0,F=F_0,\mu=\mu_{c})$, 
the true minimum is located at $(F\approx 0.8F_0,\ \Delta\approx 0.8\Delta_0)$. 
This state may be interpreted as a state in a mixed phase, in the sense that 
this phase is intermediate between a pure spin polarized phase and a pure 2SC one. 
Further, in the case (d) $\mu=0.45$ GeV, the true minimum is located at $(F\approx 0.8F_0,\ \Delta\approx 0.7\Delta_0)$. 
In the cases (e) and (f), namely $\mu\gsim 0.46$ GeV, the point $(F=F_0, \Delta=0)$ becomes a true minimum and 
the spin polarized phase is realized. 
However, in the case (e), two minima appear. 
As is seen in the case (e), it is most likely that a jump occurs from $(F=F_0, \Delta=\Delta_0)$ 
to $(F=F_0, \Delta=0)$. 
Thus, starting from 2SC phase, the second order phase transition starts at the onset of the 
spin polarization $F=F_0$, and finally it seems that the transition occurs from intermediate phase to 
spin polarized phase. 
%Thus, the transition occurs continuously from the 2SC phase with $(F=0, \Delta=\Delta_0)$ to 
%the spin polarized phase with $(F=F_0, \Delta=0)$. 
%The phase transition may be of the second order. 

%%%%%%%%%%%%%%%%%%%%%%%%%%%%%%%%%%%%%%%%%%%%%%%%%%%%%%%%%%%%%%%%%%%%%%
%\begin{figure}[t]
%\begin{center}
%\includegraphics[height=5.5cm]{PvsMu_800.eps}
%\caption{The same as Fig.1 except for three-momentum cutoff, $\Lambda=0.8$ GeV. 
%}
%\label{fig:3}
%\end{center}
%\end{figure}
%%%%%%%%%%%%%%%%%%%%%%%%%%%%%%%%%%%%%%%%%%%%%%%%%%%%%%%%%%%%%%%%%%%%%%%%

%%%%%%%%%%%%%%%%%%%%%%%%%%%%%%%%%%%%%%%%%%%%%%%%%%%%%%%%%%%%%%%%%%%%%%
%\begin{figure}[b]
%\begin{center}
%\includegraphics[height=5.5cm]{PvsRho_800.eps}
%\caption{The same as Fig.2 except for three-momentum cutoff $\Lambda=0.8$ GeV. 
%}
%\label{fig:4}
%\end{center}
%\end{figure}
%%%%%%%%%%%%%%%%%%%%%%%%%%%%%%%%%%%%%%%%%%%%%%%%%%%%%%%%%%%%%%%%%%%%%%%%

\section{Summary and concluding remarks}

In this paper, it was shown within the mean field approximation that the quark spin polarized phase may be realized after 
the two-flavor color superconducting phase and the mixed phase as the baryon density increases. 
We have first calculated the pressure of the two phases separately, namely, in the 2SC ($\Delta\neq 0$ and $F=0$) and the quark spin polarized 
($\Delta=0$ and $F\neq 0)$ phases. 
As a result, at a certain lower density region, the 2SC phase is realized. 
However, at rather high density region, the quark spin polarized phase is realized. 
If the large $F$ expansion is carried out in Eq.(\ref{35}), 
then the gap equation for $\Delta$ is expressed approximately as 
\beq\label{44}
\Delta\left[
\int\frac{d^3{\mib p}}{(2\pi)^3}\left(
1+\frac{2\mu\sqrt{p_1^2+p_2^2}-3\Delta^2}{2F^2}\right)-\frac{F}{4G_c}\right]\approx 0 \ .
\eeq
For large $F$, the above equation has only the trivial solution, $\Delta=0$. 
Thus, it is expected actually that the quark spin polarized phase with $\Delta=0$ and $F\neq 0$ 
is realized in the high baryon density region with large $F$. 
The situation is the same even if another parameter is reasonably adopted. 
For example, when we adopt $G=10$ GeV${}^{-2}$ which is one-half compared with one used in this paper, 
%the case $\Phi(F_0,\Delta=0,\mu)=\Phi(F=0,\Delta_0,\mu)$ occurs at $\mu=0.605$ GeV, while 
%the same situation occurs at $\mu=0.442$ GeV with $G=20$ GeV${}^{-2}$. 
%The behavior of phase transition may be similar, 
the behavior of the phase transition is similar in the case of $G=20$ GeV${}^{-2}$ except for 
the value of chemical potential $\mu=\mu_c=0.605$ GeV where $\Phi(\Delta=\Delta_0,F=0,\mu_c)=\Phi(\Delta=0,F=F_0,\mu_c)$ is satisfied. 
But since $\mu_c=0.605$ GeV is close to three momentum cutoff $\Lambda$, 
it may be impossible to take conclusions about the phase transition definitely.

It might be thought that, as a result, the spin polarized (SP) phase may be more favorable than the two-flavor 
color superconducting (2SC) phase at high density.
It should be noted that the thermodynamic potential can be reexpressed as 
$\Phi=-\int_0^{\mu}d\mu\ N$, which is proportional to $-\mu\times$ (volume surrounded by the Fermi surface). 
In the quark matter with $F=0$ at low baryon density or small quark chemical potential, 
the thermodynamic potential depends on $\mu^4\ (=\mu\times \mu^3)$, namely 
$\Phi\propto -\mu^4$, in which $\mu^3$ is nothing but the volume 
surrounded by the Fermi surface because 
the shape of the Fermi surface is sphere. 
However, at very high baryon density or large quark chemical potential, 
in which the relation $F>\mu$ is satisfied, the shape of Fermi surface becomes torus, 
namely, 
$(\sqrt{p_1^2+p_2^2}-F)^2+p_3^2=\mu^2$. 
Here, the volume of torus is $2\pi^2\mu^2\times F$, where $\mu$ is smaller radius 
and $F$ is the other larger radius of torus. 
In the case $F>\mu$, namely full polarization case, 
from Eq.(\ref{38}), $F=(G/(2\pi))\times \mu^3$ gives the minimum of the thermodynamic potential. 
As a result, the thermodynamic potential is proportional to $-\mu\times$ (volume surrounded by the Fermi surface) 
$\propto -\mu\times \mu^2\times F \propto -\mu^6$ as was shown in our previous paper \cite{1}. 
Thus, at high baryon density or large quark chemical potential, the spin polarized phase is favored.  
It might be concluded that the mechanism which makes SP phase more favored than 2SC phase at high densities is the distortion 
effect of the Fermi surface.

For three-momentum cutoff $\Lambda=0.631$ GeV, it is sufficient to take into account the tail of the occupation number 
since the values of the chemical potential on the phase transition point from 2SC phase to the mixed phase and 
from the mixed phase to the spin polarized phase are about 0.407 GeV and 0.46 GeV, respectively. 
In general in the system with finite chemical potential, a chemical-potential dependent cutoff may be used \cite{4}. 
However, for the sake of comparison, if a rather large fixed value of the three-momentum cutoff, $\Lambda=0.8$ GeV, is adopted 
where the tail of occupation number is more fully taken into account, 
the critical chemical potential from 2SC phase to the mixed phase is unchanged. 
Above $\mu=\mu_c=0.491$ GeV, the state with $(F=F_0, \Delta=0)$ is favored against the 
state with $(F=0, \Delta=\Delta_0)$. 
With this larger fixed three-momentum cutoff, it is concluded that, 
up to $\rho_B=7.42 \rho_0$, the color superconducting phase is realized. 
The use of large three-momentum cutoff leads to the shift of the phase transition density, that is, 
the phase transition density is higher with larger three-momentum cutoff.   
However, if a larger $\Lambda$ is used, it is necessary to re-adjust the coupling constants.

Secondly, the quark-pairing gap $\Delta$ and the spin polarization $F$ were considered simultaneously. 
Under the present treatment, the behavior of phase transition from 2SC phase with $\Delta\neq 0$ and $F=0$ 
to the quark spin polarized phase with $\Delta=0$ and $F\neq 0$ may be clear 
and the transition occurs passing through the state with $(F\neq 0, \Delta \neq 0)$ which can be interpreted as the mixed phase.

In a future work, it is interesting and important to investigate the phase equilibrium at finite 
temperature and in the presence of an external magnetic field. 
Also, matter in beta equilibrium will be analyzed, while symmetric matter is considered in the present paper.  
Also, it is important to investigate the interplay between color-flavor locked (CFL) phase and the spin polarized phase at 
higher density region in quark matter. 
Further, it may be interesting to investigate the interplay between two quark spin polarized phases, namely, 
the quark spin polarized phase originated from the tensor-type four-point interaction and 
that originated from the pseudovector-type four-point interaction between quarks. 
In this paper, the mean field approximation for the established BCS theory is 
adopted under which the thermodynamic potential is considered in the standard manner. 
Of course, there are other approaches to investigate the phase transition, such as 
by the use of the CJT potential \cite{CJT}, the Landau potential \cite{HTF} and so on. 
It would be interesting to clarify the difference between the
standard BCS mean field approach and the CJT potential approach including
the effects of two-loop order as a general problem of the phase transition
phenomena. However, that is beyond the scope of the present paper.
These topics are left for future investigations.

\section*{Acknowledgment}

One of the authors (Y.T.) would like to express his sincere thanks to 
Professor\break
J. da Provid\^encia and Professor C. Provid\^encia, two of co-authors of this paper, 
for their warm hospitality during his visit to Coimbra in summer of 2012. 
%Two of the authors (J.P. and C.P.) acknowledge valuable
%discussions with H. Bohr and\break
%P. K. Panda.
One of the authors (Y.T.) 
is partially supported by the Grants-in-Aid of the Scientific Research 
(No.23540311) from the Ministry of Education, Culture, Sports, Science and 
Technology in Japan.


%\bibliographystyle{ptephy}
%\bibliography{sample}

\begin{thebibliography}{9}
\bibitem{FH}
K. Fukushima and T. Hatsuda, Rep. Prog. Phys. {\bf 74}, 014001 (2011).

\bibitem{Tamagaki}
T. Kunihiro, T. Muto, T. Takatsuka, R. Tamagaki and T. Tatsumi, Prog. Theor. Phys. Suppl. No.112, 1 (1993).

\bibitem{RW}
M. Alford, K. Rajagopal and F. Wilczek, Nucl. Phys. B {\bf 357}, 443 (1999). 

\bibitem{IB}
K. Iida and G. Baym, Phys. Rev. D {\bf 63}, 074018 (2001).

\bibitem{CS}
M. G. Alford, A. Schmitt, K. Rajagopal and T. Schafer, Rev. Mod. Phys. {\bf 80}, 1455 (2008) and references cited therein.

\bibitem{MP}
L. McLerran and R. D. Pisarski, Nucl. Phys. A {\bf 796}, 83 (2007).

\bibitem{Tatsumi}
T. Tatsumi, Phys. Lett. B {\bf 489}, 280 (2000).\\
T. Maruyama and T. Tatsumi, Nucl. Phys. A {\bf 693}, 710 (2001).\\
E. Nakano, T. Maruyama and T. Tatsumi, Phys. Rev. D {\bf 68}, 105001 (2003). 

\bibitem{Magnetar}
C. Kouveliotou et al., Nature {\bf 393}, 235 (1998). 

\bibitem{Medan}
S. Maedan, Prog. Theor. Phys. {\bf 118}, 729 (2007).

\bibitem{1}
Y. Tsue, J. da Provid\^encia, C. Provid\^encia and M. Yamamura, Prog. Theor. Phys. {\bf 128}, 507 (2012).

\bibitem{arXiv}
H. Bohr, P. K. Panda, C. Provid\^encia and J. da Provid\^encia, 
Int. J. Mod. Phys. E {\bf 22}, 1350019 (2013).

\bibitem{2}
H. Bohr, C. Provid\^encia and J. da Provid\^encia, Eur. Phys. J. A {\bf 41}, 355 (2009).

\bibitem{NJL}
Y. Nambu and G. Jona-Lasinio, Phys. Rev. {\bf 122}, 345 (1961), Phys. Rev {\bf 124}, 246 (1961).

\bibitem{Kitazawa}
M. Kitazawa, T. Koide, T. Kunihiro and Y. Nemoto, Prog. Theor. Phys. {\bf 108}, 929 (2002).

\bibitem{3}
H. Bohr, P. K. Panda, C. Provid\^encia and J. da Provid\^encia, Braz.\ J. Phys. {\bf 42}, 68 (2012).

\bibitem{HK}
T. Hatsuda and T. Kunihiro, Phys. Rep. {\bf 247}, 221 (1994).

\bibitem{4}
C. H. Lenzi, A. S. Schneider, C. Provid\^encia and R. M. Marinho Jr., Phys. Rev. C {\bf 82}, 015809 (2010).\\
R. Casalbuoni, R. Gatto, G. Nardulli and M. Ruggoeri, Phys. Rev. D {\bf 68}, 034024 (2003).\\
M. Baldo, G. F. Burgio, P. Castorina, S. Plumari and Z. Zappal`a, Phys. Rev. C {\bf 75}, 035804 (2007).

\bibitem{CJT}
J. M. Cornwall, R. Jackiw and E. Tombolis, Phys. Rev. D {\bf 10}, 2428 (1974).

\bibitem{HTF}
Y. Hashimoto, Y. Tsue and H. Fujii, Prog. Theor. Phys. {\bf 114}, 595 (2005).

\end{thebibliography}
%
% once the .bbl file has been generated then place the text in your article.

%\vfill\pagebreak

\vspace{-0cm}

\appendix

\section{Mean field Hamiltonian in the basis of good helicity states}

The following matrix is introduced in order to diagonalize the Hamiltonian matrix 
in Eq.(\ref{5add}) with Eq.(\ref{7}): 
\beq\label{9}
V({\mib p})
=\left(
\begin{array}{cccc}
\frac{\sqrt{\varepsilon_{\mib p}^{(+)}+e+q}}{2\sqrt{\varepsilon_{\mib p}^{(+)}}} 
& -\frac{\sqrt{\varepsilon_{\mib p}^{(-)}-e+q}}{2\sqrt{\varepsilon_{\mib p}^{(-)}}} 
& -\frac{\sqrt{\varepsilon_{\mib p}^{(+)}-e-q}}{2\sqrt{\varepsilon_{\mib p}^{(+)}}} 
& \frac{\sqrt{\varepsilon_{\mib p}^{(-)}+e-q}}{2\sqrt{\varepsilon_{\mib p}^{(-)}}} \\
\frac{\sqrt{\varepsilon_{\mib p}^{(+)}+e+q}}{2\sqrt{\varepsilon_{\mib p}^{(+)}}} 
& \frac{\sqrt{\varepsilon_{\mib p}^{(-)}-e+q}}{2\sqrt{\varepsilon_{\mib p}^{(-)}}} 
& -\frac{\sqrt{\varepsilon_{\mib p}^{(+)}-e-q}}{2\sqrt{\varepsilon_{\mib p}^{(+)}}} 
& -\frac{\sqrt{\varepsilon_{\mib p}^{(-)}+e-q}}{2\sqrt{\varepsilon_{\mib p}^{(-)}}} \\
-\frac{g}{2\sqrt{\varepsilon_{\mib p}^{(+)}(\varepsilon_{\mib p}^{(+)}+e+q)}} 
& \frac{g}{2\sqrt{\varepsilon_{\mib p}^{(-)}(\varepsilon_{\mib p}^{(-)}-e+q)}} 
& -\frac{g}{2\sqrt{\varepsilon_{\mib p}^{(+)}(\varepsilon_{\mib p}^{(+)}-e-q)}} 
& \frac{g}{2\sqrt{\varepsilon_{\mib p}^{(-)}(\varepsilon_{\mib p}^{(-)}+e-q)}} \\
\frac{g}{2\sqrt{\varepsilon_{\mib p}^{(+)}(\varepsilon_{\mib p}^{(+)}+e+q)}} 
& \frac{g}{2\sqrt{\varepsilon_{\mib p}^{(-)}(\varepsilon_{\mib p}^{(-)}-e+q)}} & 
\frac{g}{2\sqrt{\varepsilon_{\mib p}^{(+)}(\varepsilon_{\mib p}^{(+)}-e-q)}} 
& \frac{g}{2\sqrt{\varepsilon_{\mib p}^{(-)}(\varepsilon_{\mib p}^{(-)}+e-q)}}
\end{array}
\right)\ . \qquad
\eeq
Then, we can diagonalize the Hamiltonian matrix $\kappa$ as $V^{\dagger}\kappa V$. 
Namely, 
\beq\label{10}
& &H_{MF}^{SP}=\sum_{{\mib p}\tau\alpha}
\left(
\begin{array}{c}
b^{\dagger}_{{\mib p}+\tau\alpha} \\
b^{\dagger}_{{\mib p}-\tau\alpha} \\
{\tilde b}^{\dagger}_{{\mib p}+\tau\alpha} \\
{\tilde b}^{\dagger}_{{\mib p}-\tau\alpha}
\end{array}
\right)^{T}
{\rm diag}\ \kappa 
\left(
\begin{array}{c}
b_{{\mib p}+\tau\alpha} \\
b_{{\mib p}-\tau\alpha} \\
{\tilde b}_{{\mib p}+\tau\alpha} \\
{\tilde b}_{{\mib p}-\tau\alpha}
\end{array}
\right)\ , 
\eeq
where we define 
\beq\label{11}
& &{\rm diag}\ \kappa
=\left(
\begin{array}{cccc}
\varepsilon_{\mib p}^{(+)} & 0 & 0 & 0 \\
0 & \varepsilon_{\mib p}^{(-)} & 0 & 0 \\
0 & 0 & -\varepsilon_{\mib p}^{(+)} & 0 \\
0 & 0 & 0 & -\varepsilon_{\mib p}^{(-)} 
\end{array}
\right)\ , 
\nonumber\\
& &\left(
\begin{array}{c}
b_{{\mib p}+\tau\alpha} \\
b_{{\mib p}-\tau\alpha} \\
{\tilde b}_{{\mib p}+\tau\alpha} \\
{\tilde b}_{{\mib p}-\tau\alpha}
\end{array}
\right)
=V^{\dagger}({\mib p})
\left(
\begin{array}{c}
c_{{\mib p}+\tau\alpha} \\
c_{{\mib p}-\tau\alpha} \\
{\tilde c}_{{\mib p}+\tau\alpha} \\
{\tilde c}_{{\mib p}-\tau\alpha}
\end{array}
\right)\ . 
\eeq

Secondly, let us consider the total Hamiltonian matrix with quark-pair interaction term in the 
above derived basis. 
In this basis, $V_{cs}$ in (\ref{6}) is written as 
\beq\label{12}
V_{cs}
=\Delta \sum_{{\mib p}\alpha\alpha'\alpha''}
\left(
\begin{array}{c}
b^{\dagger}_{{\mib p}++\alpha} \\
b^{\dagger}_{{\mib p}-+\alpha} \\
{\tilde b}^{\dagger}_{{\mib p}++\alpha} \\
{\tilde b}^{\dagger}_{{\mib p}-+\alpha}
\end{array}
\right)^{T}
V^{\dagger}({\mib p})V(-{\mib p}) 
\left(
\begin{array}{c}
b^{\dagger}_{-{\mib p}+-\alpha'} \\
b^{\dagger}_{-{\mib p}--\alpha'} \\
{\tilde b}^{\dagger}_{-{\mib p}+-\alpha'} \\
{\tilde b}^{\dagger}_{-{\mib p}--\alpha'}
\end{array}
\right)\epsilon_{\alpha\alpha'\alpha''}\  .
\eeq
Here, if we replace ${\mib p}$ into $-{\mib p}$, then 
$q$, $e$, $g$ and $\varepsilon_{\mib p}^{(\pm)}$ in (\ref{7}) are changed or unchanged to 
$q$, $e$, $-g$ and $\varepsilon_{\mib p}^{(\pm)}$. 
Thus, we obtain 
\beq\label{13}
V^{\dagger}({\mib p})V(-{\mib p})
=\left(
\begin{array}{cccc}
\frac{q+e}{\varepsilon_{\mib p}^{(+)}} & 0 & -\frac{|g|}{\varepsilon_{\mib p}^{(+)}} & 0 \\
0 & \frac{q-e}{\varepsilon_{\mib p}^{(-)}} & 0 & -\frac{|g|}{\varepsilon_{\mib p}^{(-)}} \\
-\frac{|g|}{\varepsilon_{\mib p}^{(+)}} & 0 & -\frac{q+e}{\varepsilon_{\mib p}^{(+)}} & 0 \\
0 & -\frac{|g|}{\varepsilon_{\mib p}^{(-)}} & 0 & -\frac{q-e}{\varepsilon_{\mib p}^{(-)}}
\end{array}
\right)\ .
\eeq

Finally, let us diagonalize the above matrix $V^{\dagger}({\mib p})V(-{\mib p})$. 
We introduce the following matrix:
\beq\label{14}
W=
\left(
\begin{array}{cccc}
-\sqrt{\frac{\varepsilon_{\mib p}^{(+)}+e+q}{2\varepsilon_{\mib p}^{(+)}}} & 0 & 
\sqrt{\frac{\varepsilon_{\mib p}^{(+)}-e-q}{2\varepsilon_{\mib p}^{(+)}}} & 0 \\
0 & -\sqrt{\frac{\varepsilon_{\mib p}^{(-)}-e+q}{2\varepsilon_{\mib p}^{(-)}}} & 0
& \sqrt{\frac{\varepsilon_{\mib p}^{(-)}+e-q}{2\varepsilon_{\mib p}^{(-)}}} \\
\frac{|g|}{\sqrt{2\varepsilon_{\mib p}^{(+)}(\varepsilon_{\mib p}^{(+)}+e+q)}} & 0 
& \frac{|g|}{\sqrt{2\varepsilon_{\mib p}^{(+)}(\varepsilon_{\mib p}^{(+)}-e-q)}} & 0 \\
0 & \frac{|g|}{\sqrt{2\varepsilon_{\mib p}^{(-)}(\varepsilon_{\mib p}^{(-)}-e+q)}} & 0 
& \frac{|g|}{\sqrt{2\varepsilon_{\mib p}^{(-)}(\varepsilon_{\mib p}^{(-)}+e-q)}}  
\end{array}
\right)\ .\qquad
\eeq
Then, by using the matrix $W$, the matrix $V^{\dagger}({\mib p})V(-{\mib p})$ which appears 
in the quark-pair interaction part can be diagonalized as 
\beq\label{15}
W^{\dagger}V^{\dagger}({\mib p})V(-{\mib p})W=
\left(
\begin{array}{cccc}
1 & 0 & 0 & 0 \\
0 & 1 & 0 & 0 \\
0 & 0 & -1 & 0 \\
0 & 0 & 0 & -1
\end{array}
\right)\ . 
\eeq
By introducing new fermion operators 
$(a_{{\mib p}\eta \tau \alpha}, a^{\dagger}_{{\mib p}\eta \tau \alpha},
{\tilde a}_{{\mib p}\eta \tau \alpha}, {\tilde a}^{\dagger}_{{\mib p}\eta \tau \alpha})$ by 
\beq\label{16}
\left(
\begin{array}{c}
a_{{\mib p}+\tau\alpha}\\
a_{{\mib p}-\tau\alpha}\\
{\tilde a}_{{\mib p}+\tau\alpha}\\
{\tilde a}_{{\mib p}-\tau\alpha}
\end{array}
\right)
=W^{\dagger}
\left(
\begin{array}{c}
b_{{\mib p}+\tau\alpha}\\
b_{{\mib p}-\tau\alpha}\\
{\tilde b}_{{\mib p}+\tau\alpha}\\
{\tilde b}_{{\mib p}-\tau\alpha}
\end{array}
\right) \ , 
\eeq
the mean field Hamiltonian in which both quark spin polarization and quark-pair condensate 
are simultaneously considered can be expressed in Eq.(\ref{17}).

\section{Expectation values of bilinear operators with respect to the BCS state}

The expectation values for the BCS state are summarized as follows:
\beq\label{21}
& &X_{{\mib p}\eta}=\bra{\Psi}a^{\dagger}_{{\mib p}\eta\tau\alpha}a_{{\mib p}\eta\tau\alpha'}
\ket{\Psi} \ , \qquad ({\rm for}\quad \alpha\neq \alpha')\nonumber\\
& &N_{{\mib p}\eta}=\bra{\Psi}a^{\dagger}_{{\mib p}\eta\tau\alpha}a_{{\mib p}\eta\tau\alpha}
\ket{\Psi} \ , \nonumber\\
& &D_{{\mib p}\eta}=\bra{\Psi}a_{-{\mib p}\eta-\tau\alpha'}a_{{\mib p}\eta\tau\alpha}
\ket{\Psi} \ , \qquad ({\rm for}\quad \alpha\neq \alpha') \nonumber\\
& &P_{{\mib p}\eta}=\bra{\Psi}a_{-{\mib p}\eta-\tau\alpha}a_{{\mib p}\eta\tau\alpha}
\ket{\Psi} \ ,
\eeq
where we take $\alpha=1$ and $\alpha'=2$ for the following calculations 
because color symmetry is retained. 
We can easily calculate the above expectation values by using the relation (\ref{20}) with 
(\ref{19}). 

For $\varepsilon_{\mib p}^{(\eta)}>\mu$, the following relations are obtained: 
\beq\label{22}
& &X_{{\mib p}\eta}=-K_{{\mib p}\eta}^2+K_{{\mib p}\eta}^2(N_{{\mib p}\eta}-X_{{\mib p}\eta})
\ , \nonumber\\
& &N_{{\mib p}\eta}=2K_{{\mib p}\eta}^2-2K_{{\mib p}\eta}^2(N_{{\mib p}\eta}-X_{{\mib p}\eta})\ , 
\nonumber\\
& &D_{{\mib p}\eta}=K_{{\mib p}\eta}-3K_{{\mib p}\eta}^2D_{{\mib p}\eta}
+K_{{\mib p}\eta}^2P_{{\mib p}\eta}\ , \nonumber\\
& &P_{{\mib p}\eta}=-2K_{{\mib p}\eta}^2P_{{\mib p}\eta}\ . 
\eeq
In the same way, for $\varepsilon_{\mib p}^{(\eta)}\leq\mu$, the following relations are also obtained:
\beq\label{23}
& &X_{{\mib p}\eta}=-{\wtilde K}_{{\mib p}\eta}^2(X_{{\mib p}\eta}-N_{{\mib p}\eta})
\ , \nonumber\\
& &N_{{\mib p}\eta}=1+2{\wtilde K}_{{\mib p}\eta}^2(X_{{\mib p}\eta}-N_{{\mib p}\eta})\ , 
\nonumber\\
& &D_{{\mib p}\eta}={\wtilde K}_{{\mib p}\eta}-3{\wtilde K}_{{\mib p}\eta}^2D_{{\mib p}\eta}
-{\wtilde K}_{{\mib p}\eta}^2P_{{\mib p}\eta}\ , \nonumber\\
& &P_{{\mib p}\eta}=2{\wtilde K}_{{\mib p}\eta}^2P_{{\mib p}\eta}\ . 
\eeq
As a result, the expectation values are calculated as 
\beq
& &{\rm For}\ \varepsilon_{\mib p}^{(\eta)}>\mu\nonumber\\
& &\qquad
X_{{\mib p}\eta}
%=\bra{\Phi}a^{\dagger}_{{\mib p}\eta\tau 1}a_{{\mib p}\eta\tau 2}\ket{\Phi}
=-\frac{K_{{\mib p}\eta}^2}{1+3K_{{\mib p}\eta}^2} \ , 
%\nonumber\\
%& &
\quad
N_{{\mib p}\eta}
%=\bra{\Phi}a^{\dagger}_{{\mib p}\eta\tau 1}a_{{\mib p}\eta\tau 1}\ket{\Phi}
=\frac{2K_{{\mib p}\eta}^2}{1+3K_{{\mib p}\eta}^2} \ , 
%\nonumber\\
%& &
\quad
D_{{\mib p}\eta}
%=\bra{\Phi}a_{-{\mib p}\eta-\tau 2}a_{{\mib p}\eta\tau 1}\ket{\Phi}
=\frac{K_{{\mib p}\eta}}{1+3K_{{\mib p}\eta}^2} \ , 
%\nonumber\\
%& &
\quad
P_{{\mib p}\eta}
%=\bra{\Phi}a_{-{\mib p}\eta-\tau 1}a_{{\mib p}\eta\tau 1}\ket{\Phi}
=0\ , \quad
\label{24}\\
& &{\rm For}\ \varepsilon_{\mib p}^{(\eta)}\leq \mu\nonumber\\
& &\qquad
X_{{\mib p}\eta}
%=\bra{\Phi}a^{\dagger}_{{\mib p}\eta\tau 1}a_{{\mib p}\eta\tau 2}\ket{\Phi}
=\frac{{\wtilde K}_{{\mib p}\eta}^2}{1+3{\wtilde K}_{{\mib p}\eta}^2} \ , 
%\nonumber\\
%& &
\quad
N_{{\mib p}\eta}
%=\bra{\Phi}a^{\dagger}_{{\mib p}\eta\tau 1}a_{{\mib p}\eta\tau 1}\ket{\Phi}
=\frac{1+{\wtilde K}_{{\mib p}\eta}^2}{1+3K_{{\mib p}\eta}^2}
=1-\frac{2{\wtilde K}_{{\mib p}\eta}^2}{1+3{\wtilde K}_{{\mib p}\eta}^2} \ , 
\nonumber\\
& &
\qquad
D_{{\mib p}\eta}
%=\bra{\Phi}a_{-{\mib p}\eta-\tau 2}a_{{\mib p}\eta\tau 1}\ket{\Phi}
=\frac{{\wtilde K}_{{\mib p}\eta}}{1+3{\wtilde K}_{{\mib p}\eta}^2} \ , 
%\nonumber\\
%& &
\quad
P_{{\mib p}\eta}
%=\bra{\Phi}a_{-{\mib p}\eta-\tau 1}a_{{\mib p}\eta\tau 1}\ket{\Phi}
=0\ . 
\label{25}
\eeq

%\newpage

\section{The states with $(\Delta\neq 0, F=0)$ and $(\Delta=0,F\neq 0)$}

It may be shown that $(\Delta_{0}, F=0)$ is a local minimum of the thermodynamic potential 
$\Phi(\Delta,F,\mu)$, 
where $\Delta_{0}$ is the solution of gap equation, 
$\partial \Phi(\Delta,F=0)/\partial \Delta=0$. 
Here, the states with 
$(\Delta_{0}, F=0)$ and $(\Delta=0, F_0)$ give extrema of the thermodynamic potential, 
where $F_0$ is the solution of gap equation $\partial \Phi(\Delta=0,F)/\partial F=0$:
\beq
& &\frac{\partial \Phi(\Delta, F, \mu)}{\partial \Delta}
=3\Delta\left[-2\cdot \frac{1}{V}\sum_{{\mib p},\eta}^\Lambda 
\frac{1}{\sqrt{(\varepsilon_{\mib p}^{(\eta)}-\mu)^2+3\Delta^2}}+\frac{1}{G_c}\right]\ , 
\label{b1}\\
& &\quad
\left\{
\begin{array}{l}
\displaystyle \frac{\partial \Phi(\Delta_{0}, F=0, \mu)}{\partial \Delta}=0\ , \qquad ({\rm gap\ equation})\nonumber\\
\\
%& &\quad
\displaystyle \frac{\partial \Phi(\Delta=0, F, \mu)}{\partial \Delta}=0\ , \nonumber
\end{array}\right. \\
& &
\frac{\partial \Phi(\Delta, F, \mu)}{\partial F}=
2\cdot\frac{1}{V}\sum_{{\mib p},\eta,\varepsilon_{\mib p}\leq \mu}\left[
2-\frac{\varepsilon_{\mib p}^{(\eta)}-\mu}{\sqrt{(\varepsilon_{\mib p}^{(\eta)}-\mu)^2+3\Delta^2}}\right]
\frac{\partial\varepsilon_{\mib p}^{(\eta)}}{\partial F}\nonumber\\
& &\qquad\qquad\qquad+
2\cdot\frac{1}{V}\sum_{{\mib p},\eta,\varepsilon_{\mib p}\geq \mu}\left[
1-\frac{\varepsilon_{\mib p}^{(\eta)}-\mu}{\sqrt{(\varepsilon_{\mib p}^{(\eta)}-\mu)^2+3\Delta^2}}\right]
\frac{\partial\varepsilon_{\mib p}^{(\eta)}}{\partial F}
%\nonumber\\
%& &\qquad\qquad\qquad
+\frac{F}{G}\ , 
\label{b2}\\
& &\quad
\left\{
\begin{array}{l}
\displaystyle \frac{\partial \Phi(\Delta=0, F, \mu)}{\partial F}=0\ , \qquad ({\rm gap\ equation})\nonumber\\
\\
%& &\quad
\displaystyle \frac{\partial \Phi(\Delta, F=0, \mu)}{\partial F}=0\ , 
\end{array}
\right.
\nonumber
\eeq
where the last equality is satisfied due to 
$\partial \varepsilon_{\mib p}^{(\eta)}/\partial F|_{F=0}=\eta
\sqrt{p_1^2+p_2^2}/p$, 
$\varepsilon_{\mib p}^{(\eta)}=|{\mib p}|$ and $\eta=\pm 1$.
The condition that the states corresponding to 
$(\Delta_{0}, F=0)$ or $(\Delta=0, F_{0})$ are a local minima 
is that the eigenvalues are positive for the following stability matrix ${\cal M}(\Delta,F)$, 
which consists of the second derivatives: 
\beq\label{b3}
{\cal M}(\Delta,F)=
\left(
\begin{array}{cc}
\displaystyle \frac{\partial^2\Phi(\Delta,F,\mu)}{\partial \Delta^2} & 
\displaystyle \frac{\partial^2\Phi(\Delta,F,\mu)}{\partial F \partial \Delta} \\
\displaystyle \frac{\partial^2\Phi(\Delta,F,\mu)}{\partial \Delta \partial F} & 
\displaystyle \frac{\partial^2\Phi(\Delta,F,\mu)}{\partial F^2}
\end{array}\right)\ , 
\eeq
where
\bsub\label{b3}
\beq
& &\frac{\partial^2 \Phi(\Delta, F, \mu)}{\partial \Delta^2}
=3\left[-2\cdot \frac{1}{V}\sum_{{\mib p},\eta}^\Lambda 
\frac{(\varepsilon_{\mib p}^{(\eta)}-\mu)^2}{[(\varepsilon_{\mib p}^{(\eta)}-\mu)^2+3\Delta^2]^{3/2}}+\frac{1}{G_c}\right]\ , 
\label{b3a}\\
& &
\frac{\partial^2 \Phi(\Delta, F, \mu)}{\partial F^2}=
2\cdot\frac{1}{V}\sum_{{\mib p},\eta,\varepsilon_{\mib p}\leq \mu}\left[
-\frac{3\Delta^2}{[(\varepsilon_{\mib p}^{(\eta)}-\mu)^2+3\Delta^2]^{3/2}}\right]
\frac{(F+\eta\sqrt{p_1^2+p_2^2})^2}{\varepsilon_{\mib p}^{(\eta)2}}\nonumber\\
& &\qquad\qquad\qquad
+2\cdot\frac{1}{V}\sum_{{\mib p},\eta,\varepsilon_{\mib p}\leq \mu}\left[
2-\frac{\varepsilon_{\mib p}^{(\eta)}-\mu}{\sqrt{(\varepsilon_{\mib p}^{(\eta)}-\mu)^2+3\Delta^2}}\right]
\frac{p_3^2}{\varepsilon_{\mib p}^{(\eta)3}}\nonumber\\
& &\qquad\qquad\qquad
+
2\cdot\frac{1}{V}\sum_{{\mib p},\eta,\varepsilon_{\mib p}\geq \mu}\left[
-\frac{3\Delta^2}{[(\varepsilon_{\mib p}^{(\eta)}-\mu)^2+3\Delta^2]^{3/2}}\right]
\frac{(F+\eta\sqrt{p_1^2+p_2^2})^2}{\varepsilon_{\mib p}^{(\eta)2}}\nonumber\\
& &\qquad\qquad\qquad
+
2\cdot\frac{1}{V}\sum_{{\mib p},\eta,\varepsilon_{\mib p}\geq \mu}\left[
1-\frac{\varepsilon_{\mib p}^{(\eta)}-\mu}{\sqrt{(\varepsilon_{\mib p}^{(\eta)}-\mu)^2+3\Delta^2}}\right]
\frac{p_3^2}{\varepsilon_{\mib p}^{(\eta)3}}
%\nonumber\\
%& &\qquad\qquad\qquad
+\frac{1}{G}\ , 
\label{b3b}\\
& &\frac{\partial^2\Phi(\Delta,F,\mu)}{\partial F\partial \Delta}
=6\Delta\cdot\frac{1}{V}\sum_{{\mib p},\eta}^{\Lambda}
\frac{\varepsilon_{\mib p}^{(\eta)}-\mu}{[(\varepsilon_{\mib p}^{(\eta)}-\mu)^2+3\Delta^2]^{3/2}}
\cdot\frac{F+\eta\sqrt{p_1^2+p_2^2}}{\varepsilon_{\mib p}^{(\eta)}} \ .
\label{b3c}
\eeq
\esub
Then, if $(\Delta=\Delta_{0}, F=0)$ is a local minimum, the eigenvalues of the stability matrix 
${\cal M}(\Delta_{0},F=0)$ have to be positive. 
Here, it should be noted that ${\cal M}(\Delta_{0},0)$ is a diagonal matrix because 
$\partial^2\Phi(\Delta_{0},F=0,\mu)/\partial F\partial \Delta=0$ due to 
$\varepsilon_{\mib p}^{(\eta)}=|{\mib p}|$ and 
$\eta=\pm 1$. 
Further, by using the gap equation for $\Delta (\neq 0)$, we obtain  
\beq\label{b5}
\frac{\partial^2\Phi(\Delta_{0}, F=0,\mu)}{\partial \Delta^2}
=6\cdot \frac{1}{V}\sum_{{\mib p},\eta}^{\Lambda}
\frac{3\Delta_0^2}{[(p-\mu)^2+3\Delta_0^2]^{3/2}}>0\ . 
\eeq
%%%%%%%%%%%%%%%%%%%%%%%%%%%%%%%%%%%%%%%%%%%%%%%%%%%%%%%%%%%%%%%%%%%
\begin{table}[b]
\caption{Numerical estimations}
%%%Table caption goes here
\label{table1}
\centering
\begin{tabular}{|c|c|c|}%%%The number of columns has to be defined here
\hline
$\mu$ / GeV & $\Delta_{0}$ / GeV & $\displaystyle \frac{\partial^2\Phi(\Delta_{0},F=0,\mu)}{\partial F^2}$ / GeV$^{2}$\\ 
\hline
0.44 & 0.0375255 & 0.00774356 \\
0.44190 & 0.0377933 & 0.00749214\\
$\mu_{c}$ & & \\
0.44195 & 0.0378003 & 0.00748553\\
0.45 & 0.0388851 & 0.00644747\\
%%%% Table body
\hline
%Parameter set II & 0.800 & 20.0 & 6.6 \\
%\hline
\end{tabular}
\end{table}
%%%End of the table
As for the other diagonal matrix element, $\partial^2 \Phi(\Delta_{0},F=0,\mu)/\partial F^2$, 
analytically the positiveness is not shown. 
Instead of analytical calculation, numerical estimation is useful just before and 
after the chemical potential $\mu=\mu_{c}$ with model parameters 
$G=20$ GeV${}^{-2}$, $G_c=6.6$ GeV${}^{-2}$ and $\Lambda=0.631$ GeV used in our paper, 
where $\mu_{c}$ gives $\Phi(\Delta_0,F=0,\mu_{c})=\Phi(\Delta=0, F_0,\mu_{c})$. 
From Table \ref{table1}, it is seen that the eigenvalues of the stability matrix ${\cal M}(\Delta_{0},F=0)$ are positive 
together with (\ref{b5}). 
Thus, the state with $(\Delta=\Delta_{0}, F=0)$ is identified with a local minimum state 
of the thermodynamic potential.

Next, let us consider the state with $(\Delta=0, F_{0})$. 
The stability matrix ${\cal M}(\Delta=0, F_{0})$ is also a diagonal matrix.
We easily obtain the inequality  
\beq\label{b6}
\frac{\partial^2\Phi(\Delta=0, F_{0},\mu)}{\partial F^2}
=6\cdot \frac{1}{V}\sum_{{\mib p},\eta,\varepsilon_{\mib p}^{(\eta)}\leq \mu}
\frac{p_3^2}{\varepsilon_{\mib p}^{(\eta)3}}+\frac{1}{G} >0\ . 
\eeq
As for the other matrix element,
$\partial^2\Phi(\Delta=0,F_{0},\mu)/\partial\Delta^2$ , numerical
estimation, just before and after the point
$\mu=\mu_{c}$, may also be useful. However, the integral diverges
at $\varepsilon_{\mib p}^{(\eta)}=\mu.$ This is not surprising, in view of
a well-known ``anomaly" of the BCS theory, according to which a
perturbation expansion in powers of the coupling constant $G$ is not
valid, even if this parameter is infinitesimally small. As a result,
the thermodynamical potential cannot be expanded in powers of
$\Delta$, even if this parameter is infinitesimally small. We find
that $\partial\Phi(\Delta=0,F_{0},\mu)/\partial\Delta=0$, but the
second derivative diverges. We wish to see if, in the $\Delta$
direction, the extremum point $(\Delta=0,F_{0})$ is a maximum or a
minimum. 
Figure \ref{fig:contour} shows that the point with $(\Delta=0, F_0)$ is a 
minimum point in the $F$ direction, but is a maximum point in the $\Delta$ direction in the region 
$\mu < 0.46$ GeV. 
Thus, in this region, a point $(\Delta=0, F_0)$ is a saddle point. 
However, the case $\mu \gsim 0.46$ GeV, 
the extremum point with $(\Delta=0,F_0)$ is a minimum 
point in the $\Delta$ direction. 
Thus, it is concluded that 
the point $(\Delta=0, F_0)$ gives a true minimum 
of the thermodynamic potential $\Phi$, which results the quark spin polarized phase.

\end{document}